\documentclass[11pt,a4paper]{amsart}
\usepackage{amsmath}
\usepackage{amssymb}
\usepackage{amsfonts}
\usepackage[toc,page]{appendix}
\usepackage{hyperref}
\usepackage{color}

\def\={\ =\ }

\newcommand{\be}{\begin{equation}}
\newcommand{\ee}{\end{equation}}
\newcommand{\beq}{\begin{equation}}
\newcommand{\eeq}{\end{equation}}
\newcommand{\bea}{\begin{eqnarray}}
\newcommand{\eea}{\end{eqnarray}}

\def\ba{\begin{eqnarray}}
\def\ea{\end{eqnarray}}

\theoremstyle{plain}

\numberwithin{equation}{section}

\begin{document}
\title[]{Mordell integrals and Giveon-Kutasov duality}
\author{Georgios Giasemidis}
\author{Miguel Tierz}
\address{Countinglab LTD \& Centre for the Mathematics of Human Behaviour
(CMoHB), Department of Mathematics and Statistics, University of Reading,
RG6 6AX, Reading, U.K.}
\email{g.giasemidis@reading.ac.uk}
\address{Departamento de Matem\'{a}tica, Grupo de F\'{\i}sica Matem\'{a}%
tica, Faculdade de Ci\^{e}ncias, Universidade de Lisboa, Campo Grande, Edif%
\'{\i}cio C6, 1749-016 Lisboa, Portugal.}
\email{tierz@fc.ul.pt}
\address{Departamento de An\'{a}lisis Matem\'{a}tico, Facultad de Ciencias
Matem\'{a}ticas, Universidad Complutense de Madrid, 28040 Madrid, Spain}
\email{tierz@mat.ucm.es}
\maketitle

\begin{abstract}
We solve, for finite $N$, the matrix model of supersymmetric $U(N)$
Chern-Simons theory coupled to $N_{f}$ massive hypermultiplets of $R$-charge 
$\frac{1}{2}$, together with a Fayet-Iliopoulos term. We compute the
partition function by identifying it with a determinant of a Hankel matrix,
whose entries are parametric derivatives (of order $N_{f}-1$) of Mordell
integrals. We obtain finite Gauss sums expressions for the partition
functions. We also apply these results to obtain an exhaustive test of
Giveon-Kutasov (GK) duality in the $\mathcal{N}=3$ setting, by systematic
computation of the matrix models involved. The phase factor that arises in
the duality is then obtained explicitly.\ We give an expression
characterized by modular arithmetic (mod 4) behavior that holds for all
tested values of the parameters (checked up to $N_{f}=12$ flavours).
\end{abstract}

\section{Introduction}

The study of supersymmetric gauge theories has greatly benefited in recent
years from the development of the localization of supersymmetric gauge
theories by Pestun \cite{Pestun:2007rz} (\cite%
{Pestun:2014mja,Hosomichi:2015jta,Cremonesi:2014dva} for recent reviews).
The localization procedure manages to reduce the original functional
integral describing a quantum field theory into a much simpler matrix
integral. Thus, it enormously reduces the task of computing observables in a
supersymmetric gauge theory. However, there still remains the issue of
explicitly computing $N$ integrations, in which case one needs to employ
matrix model tools \cite{For} in order to obtain explicit expressions for
the observables of the gauge theory.

The theory we shall focus on is ${\mathcal{N}}=2$ and ${\mathcal{N}}=3$
supersymmetric $U(N)$ Chern-Simons (CS) on three-sphere, $\mathbb{S}^{3}$,
with $N_{f}$ fundamental and $N_{f}$ antifundamental chiral multiplets of
mass $m$. Indeed the partition function on $\mathbb{S}^{3}$ can be
determined by the localization techniques of \cite{Pestun:2007rz}, which
were used in the 3d case in \cite%
{Kapustin:2009kz,Kapustin:2010xq,Jafferis:2010un,Hama:2010av}. In the case
of the partition function for $U(N)$ $\mathcal{N}=2$ Chern-Simons theory at
level $k$ coupled to $N_{f}$ fundamental and $\bar{N}_{f}$ anti-fundamental
chiral multiplets of $R$-charge $q$ the matrix model is \cite{Hama:2010av}%
\footnote{%
Notice that we have changed the sign of the Chern-Simons level with respect
to that in \cite{Hama:2010av} in order to make contact with our conventions.}%
\begin{equation}
Z=\frac{1}{N!}\int d^{N}\sigma \;\prod_{j=1}^{N}e^{\mathrm{i}\pi k\sigma
_{j}^{2}}\left( s_{b=1}(\mathrm{i}-\mathrm{i} q-\sigma _{j})\right)
^{N_{f}}\left( s_{b=1}(\mathrm{i}-\mathrm{i} q+\sigma _{j})\right) ^{\bar{N}%
_{f}}\prod_{i<j}^{N}(2\sinh \pi (\sigma _{i}-\sigma _{j}))^{2},  \label{Zds}
\end{equation}%
where $s_{b=1}(\sigma )$ denotes the double sine function \cite%
{Jafferis:2010un,Hama:2010av} (and references therein). This matrix model
corresponds to the case where the matter chiral multiplets have $R$-charge $%
q $ and belong to the representation $R$ of the gauge group. The fact that
for $\mathcal{N}=3$ theories the R-symmetry is non-abelian allows us to fix
an $R$-charge which is not altered under the RG flow. In this paper, we
focus on a detailed study of the case where $q=1/2$ and $R=r\oplus \overline{%
r}$. In this case, due to the basic property of the double sine function 
\cite{Hama:2010av} 
\begin{equation}
\prod_{\rho \in r}s_{b=1}(\tfrac{\mathrm{i}}{2}-\rho _{i}\hat{\sigma}%
_{i})\cdot s_{b=1}(\tfrac{\mathrm{i}}{2}+\rho _{i}\hat{\sigma}%
_{i})~=~\prod_{\rho \in r}\frac{1}{2\cosh \pi \rho _{i}\hat{\sigma}_{i}}\,,
\label{dsid}
\end{equation}%
and setting $\bar{N}_{f}=N_{f}$, the matter contribution simplifies, leaving
the matrix model to be%
\begin{equation}
Z_{N_{f}}^{U(N)}=\frac{1}{\left( 2\pi \right) ^{N}N!}\int {d^{N}\!\mu }\frac{%
\prod_{i<j}4\sinh ^{2}(\frac{1}{2}(\mu _{i}-\mu _{j}))\ e^{-\frac{1}{2g}%
\sum_{i}\mu _{i}^{2}+\mathrm{i} \eta \sum_{i}\mu _{i}}}{\prod_{i}\left(
2\cosh (\frac{1}{2}(\mu _{i}+m))\right) ^{N_{f}}}\ ,  \label{Z}
\end{equation}%
where $g=\frac{2\pi \mathrm{i}}{k}$ with $k\in \mathbb{Z}$ the Chern-Simons
level and $\mu _{i}/2\pi $ represent the eigenvalues of the scalar field $%
\sigma $ belonging to the three dimensional %$\mathcal{N}=2$ 
vector multiplet. In \eqref{Z} the radius $\mathrm{R}$ of the three-sphere
has been set to one. It can be restored by rescaling $m\rightarrow m \mathrm{%
R}$, $\mu _{i}\rightarrow \mu _{i} \mathrm{R}$. The partition function is
periodic in imaginary shifts of the mass, $Z(m+\mathrm{i} 2\pi n)=Z(m)$, for
integer $n$. The addition of a Fayet-Iliopoulos term (FI) in the Lagrangian
adds a linear term in the potential of the matrix model \cite%
{Kapustin:2009kz,Kapustin:2010xq,Hama:2010av}. Thus $\eta $ is a real
parameter denoting the FI parameter. Notice that the variables in (\ref{Z})
are rescaled with a $2\pi $ factor with regard to those in \cite%
{Kapustin:2009kz,Kapustin:2010xq,Hama:2010av} and with regard to the ones in
(\ref{Zds}). That is, $\mu _{i}=2\pi \sigma _{i}$.

We shall focus in this work specifically on the model (\ref{Z}) but also
consider a variant of the same model, with matter content a pair of
fundamental and a pair of anti-fundamental chiral multiplets ($N_{f}$
hypermultiplets of mass $m$ and $N_{f}$ hypermultiplets of mass $-m$). The
corresponding matrix model is %in (\ref{Zm2}) below. 
\begin{equation}
\widetilde{Z}_{N_{f}}^{U(N)}=\frac{1}{\left( 2\pi \right) ^{N}N!}\int {\
d^{N}\!\mu }\frac{\prod_{i<j}4\sinh ^{2}(\frac{1}{2}(\mu _{i}-\mu _{j}))\
e^{-\frac{1}{2g}\sum_{i}\mu _{i}^{2}}}{\prod_{i}\left( 4\cosh (\frac{1}{2}%
(\mu _{i}+m))\cosh (\frac{1}{2}(\mu _{i}-m))\right) ^{N_{f}}}\,,  \label{Zm2}
\end{equation}%
which was previously studied for large $N$ in \cite{BR} and for finite $N$
and $N_{f}=1$ in \cite{Russo:2014bda}. In this paper we consider both
models, as spelled out in detail in the next Section.

In \cite{Russo:2014bda}, the approach is to express 
%such matrix model, (\ref{Zm2}) below, 
the matrix model \eqref{Zm2} for $N_{f}=1,$ as a Hankel determinant whose
entries are (combinations of) Mordell integrals \cite{Mordell}%
\begin{equation}
I(l,m)=\int_{-\infty }^{\infty }d\mu \frac{e^{(l+1)\mu +m}}{1+e^{\mu +m}}%
e^{-\mu ^{2}/2g},  \label{imor}
\end{equation}%
where $l\in 
%TCIMACRO{\U{2102} }%
%BeginExpansion
\mathbb{C}
%EndExpansion
$ \footnote{%
That will be the case if there is a FI parameter, see below. In simpler
settings, such as in \cite{Russo:2014bda} and also below, it may only be an
integer.}. This integral, $I(l,m)$, was computed by Mordell \cite{Mordell}
for general parameters. In general, it is given in terms of infinite sums of
the theta-function type. However, in specific cases it assumes the form of a
Gauss's finite sum \cite{Mordell,Russo:2014bda}. These specific cases
precisely contain the one which is physically relevant: $g=2\pi \mathrm{i}/k$
with $k\in \mathbb{Z}$. Exactly the same method can be applied to (\ref{Z})
and, as a matter of fact, it is simpler in that case since the
identification with the Mordell integral is more direct, as we shall see
below.

The main difference between \cite{Russo:2014bda}, where analytical results
for the case $N_{f\text{ }}=1$ were given using Mordell integrals, and this
work, can be succinctly summarized by 
%noting that we will substitute (\ref{imor}) by 
substituting \eqref{imor} for 
\begin{equation}
J(l,m)=\int_{-\infty }^{\infty }d\mu \frac{e^{(l+1)\mu +m}}{\left( 1+e^{\mu
+m}\right) ^{N_{f}}}e^{-\mu ^{2}/2g},  \label{gen}
\end{equation}%
as the main building block in all computations. This allows us to do the
same computations as in \cite{Russo:2014bda}, but also for higher flavour
cases $N_{f\text{ }}\geq 1$ and for both 
%(\ref{Z}) above and (\ref{Zm2}) below. 
\eqref{Z} and \eqref{Zm2} above. The first task will then be to establish
all the above-mentioned properties for \eqref{gen}, following from those of (%
\ref{imor}). We will achieve this by using (\ref{imor}) and systematically
differentiating under the integral sign, establishing also a recurrence
relationship between the different derivatives. At this stage, it is worth
mentioning that Mordell integrals have been applied before in problems of
theoretical physics, in particular, in the study of superconformal algebras 
\cite{Eguchi:1988af,Eguchi:2010cb}. In number theory, they have become
especially prominent in the last decade, after \cite{Mock}, due to their
intimate relationship with Mock theta functions and also due to their
modular properties. For example, in \cite{FB} we find an analysis of roughly
the same generalization of the integral, namely (\ref{gen}), but not in the
physical setting required to study (\ref{Z}) \footnote{%
The analysis in \cite{FB} focusses on real values of the parameters in the
exponential in the context of a heat-kernel expansion, which is not the
physical setting needed in the analysis of our matrix models.}.

The main use of the formulas derived, apart from the computation of the
partition functions (\ref{Z}) %(and (\ref{Zm2}) below), 
and \eqref{Zm2}, is as a tool to further analyze a Seiberg-like duality in a
3d theory \cite{Giveon:2008zn,Aharony:1997gp}. The two main types of
Seiberg-like dualities in 3d are:

\begin{enumerate}
\item Aharony duality \cite{Aharony:1997gp} for three dimensions, which
holds for Chern-Simons level $k=0$ and is characterized by an unusual
coupling between electric and magnetic monopoles.

\item Giveon--Kutasov duality \cite{Giveon:2008zn}, which applies to
theories with any Chern-Simons level and resembles a 4d Seiberg duality for
theories with fundamental matter or an adjoint field. This allows for
compact expressions for the observables of the Chern-Simons-matter theory as
was shown in \cite{Russo:2014bda} and is discussed here as well.
\end{enumerate}

These two dualities can be related by starting with the Aharony duality and
adding masses and generating Chern-Simons terms, leading to the
Giveon--Kutasov duality. The reverse renormalization group flow from
Giveon--Kutasov duality in the UV to Aharony duality in the IR has been
studied in \cite{Intriligator:2013lca}. We shall focus here on the
Giveon-Kutasov duality, which implies for the partition function \cite%
{Kapustin:2010mh}%
\begin{equation}
Z_{N_{f},k}^{U(N_{c})}\left( \eta \right) =e^{\mathrm{sgn}(k)\pi \mathrm{i}%
\left( c_{\left\vert k\right\vert ,N_{f}}-\eta ^{2}\right)
}Z_{N_{f},-k}^{U(\left\vert k\right\vert +N_{f}-N_{c})}\left( -\eta \right) ,
\label{GK}
\end{equation}%
where the l.h.s. denotes the partition function of a theory with $N_{c}$
colors, $N_{f}$ %fundamental chiral multiplets, 
hypermultiplets, Chern-Simons level $k$, and a Fayet-Iliopoulos term $\eta $%
. The term $c_{\left\vert k\right\vert ,N_{f}}$ is a phase, which is a
quadratic polynomial in $k$. The principal result is an explicit expression,
characterized by modular arithmetic (mod 4) behavior, for the phase factor
in (\ref{GK}). Previous, conjectured results, for this phase factor can be
found in \cite{Kapustin:2010mh,Willett:2011gp,Kapustin:2013hpk}.

The paper is organized as follows. In the next Section we use the
determinant formulation of \cite{Russo:2014bda}, applying it to \eqref{Z}
while also extending it to the case $N_{f}>1$. This extension is based on
explicit finite Gauss sums expressions for the generalized Mordell integral (%
\ref{gen}) that we obtain. In Section 3, we use such analytical results,
together with their practical implementation in \textit{Mathematica}, to
obtain exact analytical expressions for both (\ref{Z}) and (\ref{Zm2}) for a
number of values of $(N,N_{f},k)$ which leads also to discuss some physical
interpretations in terms of symmetry protected phases \cite%
{Chen:2011pg,Witten:2015aba} and mathematical features like a complex
conjugacy property under the transformation $k\rightarrow -k$.

Finally, in the last Section, we use the formalism developed to perform an
exhaustive test of Giveon-Kutasov duality in the $\mathcal{N}=3$ setting, by
explicit and systematic computation of the matrix model (\ref{Z}) on both
sides of (\ref{GK}). We propose an explicit expression of the phase factor
in (\ref{GK}), which is different from previous expressions in the
literature \cite{Kapustin:2010mh,Willett:2011gp,Kapustin:2013hpk} and that
we have tested to hold for a large range of the parameter space, going up to 
$12$ flavours.

\section{Parametric derivatives of Mordell integrals for the arbitrary $%
N_{f} $ setting}

\label{NfGeneric}

\bigskip Let us develop the method based on generalized Mordell integrals in
order to compute (\ref{Z}) for higher flavour $N_{f}>1$, therefore extending
the methodology and results in \cite{Russo:2014bda}. It is enough to
directly consider the derivatives of a single Mordell integral. More
precisely, by making the change of variables 
\begin{equation}
z_{i}=c\ e^{\mu _{i}}\ ,\qquad c=e^{g\left( N-\frac{N_{f}}{2}\right) }\ ,
\label{changeofvariable1}
\end{equation}%
we may write (\ref{Z}) in the form 
\begin{equation}
Z_{N_{f}}^{U(N)}=\frac{e^{-\frac{gN}{2}\left( N-\frac{N_{f}}{2}\right)
\left( N+\frac{N_{f}}{2}+2\mathrm{i}\eta \right) }}{\left( 2\pi \right)
^{N}N!}\int_{\left[ 0,\infty \right) ^{N}}{d^{N}z}\
\prod_{i<j}(z_{i}-z_{j})^{2}\frac{e^{-\frac{1}{2g}\sum_{i}(\ln z_{i})^{2}+%
\mathrm{i}\eta \sum_{i}\ln z_{i}}}{\prod_{i}\left( 1+\frac{z_{i}e^{m}}{c}%
\right) ^{N_{f}}}\,.  \label{Z-1}
\end{equation}%
Its determinantal formulation becomes 
\begin{equation*}
Z_{N_{f}}^{U(N)}=\frac{e^{\frac{NN_{f}m}{2}}c^{-\frac{N}{2}\left( N+\frac{%
N_{f}}{2}+2\mathrm{i}\eta \right) }}{(2\pi )^{N}}\ \det \left( \left(
g_{i},g_{j}\right) \right) _{i,j=0}^{N-1}\,,
\end{equation*}%
where the matrix elements are given by%
\begin{equation}
\left( g_{i},g_{j}\right) \,\equiv c^{i+j+1+\mathrm{i}\eta }e^{-\frac{1}{2g}%
\left( \ln c\right) ^{2}}\int_{-\infty }^{\infty }d\mu \frac{e^{\mu \left(
i+j+1-N+\frac{N_{f}}{2}+\mathrm{i}\eta \right) }}{\left( 1+e^{\mu +m}\right)
^{N_{f}}}e^{-\frac{1}{2g}\mu ^{2}}\,.  \label{gigj}
\end{equation}%
The matrix elements may also be written in terms of a Mordell integral as
follows%
\begin{equation*}
\left( g_{i},g_{j}\right) \,=c^{i+j+1+\mathrm{i}\eta }e^{-\frac{1}{2g}\left(
\ln c\right) ^{2}}\frac{(-1)^{N_{f}-1}e^{-m}}{(N_{f}-1)!}\frac{d}{dm}\left(
e^{-m}\frac{d}{dm}\left( e^{-m}\frac{d}{dm}\left( e^{-m}\frac{d}{dm}\cdots 
\frac{d}{dm}\left( e^{-m}I(l,m)\right) \right) \right) \right) \,,
\end{equation*}%
where the derivative has to be applied $N_{f}-1$ times and $l=i+j+1-N-\frac{%
N_{f}}{2}+\mathrm{i}\eta $. %More succinctly, we have%
Alternatively, one may exploit the relation%
\begin{equation}
\int_{-\infty }^{\infty }d\mu \frac{e^{(l+N_{f})\mu }e^{-\mu ^{2}/2g}}{%
(1+e^{\mu +m})^{N_{f}}}=\frac{(-1)^{N_{f}}e^{-mN_{f}}}{(N_{f}-1)!}%
\sum_{n=0}^{N_{f}-1}C_{N_{f}-1,n}I^{(n)}(l,m)
\end{equation}%
and hence express \eqref{gigj} as follows%
\begin{equation*}
\left( g_{i},g_{j}\right) =c^{i+j+1+\mathrm{i}\eta }e^{-\frac{1}{2g}\left(
\ln c\right) ^{2}}\frac{(-1)^{N_{f}}}{(N_{f}-1)!}e^{-mN_{f}}%
\sum_{n=0}^{N_{f}-1}C_{N_{f}-1,n}I^{(n)}(l,m)\,,
\end{equation*}%
%
%
%
%
%
%
%
%
%where the index $^{(n)}$ 
where $I^{(n)}$ stands for the $n$-th derivative of $I(l,m)$ with respect to 
$m$ and the coefficients $C_{p,q}$ satisfy 
\begin{equation}
C_{p,q}=%
\begin{cases}
-p\ C_{p-1,q}+C_{p-1,q-1}, & p>q>0, \\ 
(-1)^{p+1}p!, & p>q=0, \\ 
-1, & p=q.%
\end{cases}
\label{Cpq}
\end{equation}%
As it was shown in \cite{Russo:2014bda}, \eqref{imor} is proportional to the
Mordell integral%
\begin{equation}
I(l,m)=2\pi e^{-ml+\mathrm{i} km^{2}/(4\pi )}\int_{-\infty }^{\infty }dt%
\frac{e^{\mathrm{i}\pi kt^{2}+2\pi t(l+1)-\mathrm{i} tkm}}{e^{2\pi t}+1}
\end{equation}%
and has the following explicit expression \cite{Russo:2014bda} 
\begin{equation}
I(l,m)=2\pi 
\begin{cases}
e^{-\mathrm{i}\pi (l+\frac{k}{4})}\ e^{-m(l+\frac{k}{2})+\frac{\mathrm{i}
km^{2}}{4\pi }}G_{+}\left( k,1,-l-1+\mathrm{i}\frac{km}{2\pi }-\frac{k}{2}%
\right) , & k>0, \\ 
e^{\mathrm{i}\pi (l-\frac{k}{4})}\ e^{-m(l-\frac{k}{2})+\frac{\mathrm{i}
km^{2}}{4\pi }}G_{-}\left( -k,1,-l-1+\mathrm{i}\frac{km}{2\pi }+\frac{k}{2}%
\right) , & k<0,%
\end{cases}%
\ ,  \label{mart}
\end{equation}%
with 
\begin{eqnarray}
&&G_{+}\left( k,1,-l-1+\mathrm{i}\frac{km}{2\pi }-\frac{k}{2}\right) =\frac{1%
}{e^{-2\pi \mathrm{i} l-km}-1}\left( -\sqrt{\frac{\mathrm{i}}{k}}\
\sum_{r=1}^{k}e^{\frac{\mathrm{i}\pi }{k}(r-l-1-\frac{k}{2}+\mathrm{i}\frac{%
km}{2\pi })^{2}}+\mathrm{i}\right) ,  \label{mert} \\
&&G_{-}\left( -k,1,-l-1+\mathrm{i}\frac{km}{2\pi }+\frac{k}{2}\right) =\frac{%
1}{1-e^{2\pi \mathrm{i} l+km}}\left( \sqrt{\frac{\mathrm{i}}{k}}e^{2\pi 
\mathrm{i} l+km}\sum_{r=1}^{-k}e^{\mathrm{i}\frac{\pi }{k}\left( r+l-\frac{k%
}{2}-\mathrm{i}\frac{km}{2\pi }\right) ^{2}}+\mathrm{i}\right) .
\end{eqnarray}%
Notice that, if $l\in \mathbb{Z}$, the denominator in, say, $G_{+}(a,b,x)$
could vanish in principle. However, a Gauss's sum identity \cite[Eq. 2. 28]%
{Russo:2014bda} prevents the partition function to diverge. We will see that
for $N_{f}$ odd we get $l\in \mathbb{Z}/2$ (if we set the FI parameter to
zero); in that case the identity does not apply, but the factor $e^{-2\pi%
\mathrm{i} l}\neq 1$ implies that the denominator does not vanish. Notice
that the prefactor of $G_{+}$ in \cite{Russo:2014bda} is slightly different.
This is due to the fact that, there, we discussed the model with $2N_{f}$ \
hypermultiplets, 
%which we also study here, in Section \ref{2Nf_Hypermultiplets} below, 
which we study later in Section \ref{2Nf_Hypermultiplets} for $N_{f}>1$.

The first derivative of the Mordell integral \eqref{mart} takes the form 
\begin{equation*}
I^{\prime }(l ,m)=%
\begin{cases}
e^{-\mathrm{i}\pi (l +\frac{k}{4})}\ e^{-m(l +\frac{k}{2})+\frac{\mathrm{i}
km^{2}}{4\pi }}\Bigg((\mathrm{i} km-(k+2 l )\pi )G_{+}\left( k,1,-l -1+%
\mathrm{i}\frac{km}{2\pi }-\frac{k}{2}\right) &  \\ 
\qquad +2\pi G_{+}^{\prime }\left( k,1,-l -1+\mathrm{i}\frac{km}{2\pi }-%
\frac{k}{2}\right) \Bigg), & k>0, \\ 
e^{\mathrm{i}\pi (l -\frac{k}{4})}\ e^{-m(l -\frac{k}{2})+\frac{\mathrm{i}
km^{2}}{4\pi }}\Bigg((\mathrm{i} km+(k-2 l )\pi )G_{-}\left( -k,1,-l -1+%
\mathrm{i}\frac{km}{2\pi }+\frac{k}{2}\right) &  \\ 
\qquad +2\pi G_{-}^{\prime }\left( -k,1,-l -1+\mathrm{i}\frac{km}{2\pi }+%
\frac{k}{2}\right) \Bigg), & k<0,%
\end{cases}%
\end{equation*}%
where we also have the Gauss sums 
\begin{eqnarray}
G_{+}^{\prime }\left( k,1,-l -1+\mathrm{i}\frac{km}{2\pi }-\frac{k}{2}%
\right) =\frac{e^{km+2\mathrm{i} l \pi }}{\left( -1+e^{km+2\mathrm{i} l \pi
}\right) ^{2}}\Bigg(\sqrt{\frac{\mathrm{i}}{k}}\Bigg(-k\sum_{r=1}^{k}e^{%
\frac{\mathrm{i}\pi }{k}\left( -1-\frac{k}{2}- l +\frac{\mathrm{i} km}{2\pi }%
+r\right) ^{2}} +  \notag \\
\left( e^{km+2\mathrm{i} l \pi }-1\right) \sum_{r=1}^{k}e^{\frac{\mathrm{i}%
\pi }{k}\left( -1-\frac{k}{2}-l +\frac{\mathrm{i} km}{2\pi }+r\right)
^{2}}\left( 1+\frac{k}{2}+l -\frac{\mathrm{i} km}{2\pi }-r\right) \Bigg)+%
\mathrm{i} k\Bigg ),  \label{mert2} \\
G_{-}^{\prime }\left( -k,1,-l -1+\mathrm{i}\frac{km}{2\pi }+\frac{k}{2}%
\right) =\frac{e^{km+2\mathrm{i} l \pi }}{\left( -1+e^{km+2\mathrm{i} l \pi
}\right) ^{2}}\Bigg(\sqrt{\frac{\mathrm{i}}{k}}\Bigg(k\sum_{r=1}^{-k}e^{%
\frac{\mathrm{i}\pi }{k}\left( -\frac{k}{2}+l -\frac{\mathrm{i} km}{2\pi }%
+r\right) ^{2}} -  \notag \\
\left( -1+e^{km+2\mathrm{i} l\pi }\right) \sum_{r=1}^{-k}e^{\frac{\mathrm{i}%
\pi }{k}\left( -\frac{k}{2}+l -\frac{\mathrm{i} km}{2\pi }+r\right)
^{2}}\left( -\frac{k}{2}+l -\frac{\mathrm{i} km}{2\pi }+r\right) \Bigg)+%
\mathrm{i} k\Bigg).
\end{eqnarray}

\subsection{The theory with $2N_{f}$ Hypermultiplets}

\label{2Nf_Hypermultiplets} We now consider the case of $N_{f}$
hypermultiplets with vector mass $m$ and $N_{f}$ hypermultiplets with vector
mass $-m$. This theory was analyzed in detail in \cite{Russo:2014bda} for $%
N_{f}=1$ and previously in \cite{BR}. One of its distinctive features is the
existence of third order phase transitions in a certain double scaling limit
which involves a decompactification limit, in which the radius of $\mathbb{S}%
^{3}$ is taken to infinity \cite{BR,Russo:2014bda}. The partition function,
expressed as a matrix integral, is%
\begin{equation*}
\widetilde{Z}_{N_{f}}^{U(N)}=\frac{1}{\left( 2\pi \right) ^{N}N!}\int {%
d^{N}\!\mu }\frac{\prod_{i<j}4\sinh ^{2}(\frac{1}{2}(\mu _{i}-\mu _{j}))\
e^{-\frac{1}{2g}\sum_{i}\mu _{i}^{2}}}{\prod_{i}\left( 4\cosh (\frac{1}{2}%
(\mu _{i}+m))\cosh (\frac{1}{2}(\mu _{i}-m))\right) ^{N_{f}}}\,,
\end{equation*}%
where for simplicity (and to compare with \cite{Russo:2014bda}) we have not
included the FI term. The simple change of variables is \cite%
{BR,Russo:2014bda}%
\begin{equation}
z_{i}=ce^{\mu _{i}}\ ,\qquad c=e^{g(N-N_{f})}\ ,  \label{changeofvariable2}
\end{equation}%
which recasts the integral in the form%
\begin{equation}
\widetilde{Z}_{N_{f}}^{U(N)}=\frac{e^{-\frac{gN}{2}%
(N^{2}-N_{f}^{2})}e^{2gNN_{f}}}{\left( 2\pi \right) ^{N}N!}\int_{\left[
0,\infty \right) ^{N}}{d^{N}z}\ \prod_{i<j}(z_{i}-z_{j})^{2}\frac{e^{-\frac{1%
}{2g}\sum_{i}(\ln z_{i})^{2}}}{\prod_{i}\left( ce^{-m}+z_{i}\right)
^{N_{f}}\left( ce^{m}+z_{i}\right) ^{N_{f}}}\,.  \label{Z-2}
\end{equation}%
As in \cite{Russo:2014bda}, the partition function can be written as%
\begin{equation}
\widetilde{Z}_{N_{f}}^{U(N)}=N!e^{-\frac{gN}{2}(N^{2}-N_{f}^{2})}\det \left(
\left( f_{i},f_{j}\right) \right) _{i,j=0}^{N-1}  \label{ZNf}
\end{equation}%
where the functions $f_{i}$ have the form%
\begin{equation}
(f_{i},f_{j})=c^{i+j+1}e^{-(\ln c)^{2}/2g}\int_{-\infty }^{+\infty }d\mu 
\frac{e^{i+j+1+N_{f}-N}}{(1+e^{\mu +m})^{N_{f}}(1+e^{\mu -m})^{N_{f}}}%
e^{-\mu ^{2}/2g}.  \label{fifj}
\end{equation}%
%
%
%
%
%
%
%
%
%
%
%
%
%
%
%
%
%
%
%with $c=e^{g(N-N_{f})}$. 
The objective is to compute \eqref{fifj} for generic $N_{f}$ \ and then the
partition function follows from \eqref{ZNf}. First we expand the denominator
in \eqref{fifj} using the identity 
\begin{eqnarray}
\frac{1}{(1+ax)^{n}(1+bx)^{n}} &=&\frac{1}{(a-b)^{n}}\sum_{s=0}^{n-1}\binom{%
n+s-1}{s}\left( \frac{ab}{a-b}\right) ^{s}\Bigg((-1)^{s}\left( \frac{a}{1+ax}%
\right) ^{n-s}  \notag \\
&&+(-1)^{n}\left( \frac{b}{1+bx}\right) ^{n-s}\Bigg)  \label{denominator}
\end{eqnarray}%
This is a generalization for $n>1$ of the identity used in \cite%
{Russo:2014bda}. Using the summation form of \eqref{denominator} and setting 
$\ell =i+j+1-N$, we are able to write \eqref{fifj} as 
\begin{eqnarray}
(f_{i},f_{j}) &=&\frac{c^{i+j+1}e^{-(\ln c)^{2}/2g}}{2^{N_{f}}(\sinh
m)^{N_{f}}}\sum_{s=0}^{N_{f}-1}\binom{N_{f}+s-1}{s}\frac{1}{(2\sinh m)^{s}} 
\notag \\
&&\times \Bigg((-1)^{s}e^{m(N_{f}-s-1)}\int_{-\infty }^{\infty }d\mu \frac{%
e^{\mu (\ell +N_{f})+m}e^{-\mu ^{2}/2g}}{(1+e^{\mu +m})^{N_{f}-s}}  \notag \\
&&+(-1)^{N_{f}}e^{-m(N_{f}-s-1)}\int_{-\infty }^{\infty }d\mu \frac{e^{\mu
(\ell +N_{f})-m}e^{-\mu ^{2}/2g}}{(1+e^{\mu -m})^{N_{f}-s}}\Bigg).
\label{fifj_2}
\end{eqnarray}%
The integrals in \eqref{fifj_2} can be expressed in terms of a Mordell
integral \eqref{imor}. %In particular we use the integral 
%\begin{equation}
%I(l,m)=\int_{-\infty }^{\infty }d\mu \frac{e^{(\ell+1)\mu +m}}{1+e^{\mu +m}}%
%e^{-\mu ^{2}/2g}.  \label{MordelI}
%\end{equation}%
Specifically, we note that 
\begin{eqnarray}
&&\frac{d}{dm}\left( e^{-m}\frac{d}{dm}\left( e^{-m}\frac{d}{dm}\cdots \frac{%
d}{dm}\left( e^{-m}I(\ell +s,m)\right) \right) \right)  \notag \\
&&\qquad =(-1)^{N_{f}-s-1}(N_{f}-s-1)!\int_{-\infty }^{\infty }\frac{e^{\mu
(\ell +N_{f})+m}e^{-\mu ^{2}/2g}}{(1+e^{\mu +m})^{N_{f}-s}}, \\
&&\frac{d}{dm}\left( e^{m}\frac{d}{dm}\left( e^{m}\frac{d}{dm}\cdots \frac{d%
}{dm}\left( e^{m}I(\ell +s,-m)\right) \right) \right)  \notag \\
&&\qquad =(N_{f}-s-1)!\int_{-\infty }^{\infty }\frac{e^{\mu (\ell
+N_{f})-m}e^{-\mu ^{2}/2g}}{(1+e^{\mu -m})^{N_{f}-s}},
\end{eqnarray}%
where the derivative on the left hand side of the above expressions is
applied $N_{f}-s-1$ times. %Therefore%
%\begin{eqnarray}
%(f_{i},f_{j}) &=&\frac{c^{i+j+1}e^{-(\ln c)^{2}/2g}}{2^{N_{f}}(\sinh
%m)^{N_{f}}}\sum_{s=0}^{N_{f}-1}\binom{N_{f}+s-1}{s}\frac{1}{(2\sinh m)^{s}} 
%\notag \\
%&&\times \Bigg(\frac{(-1)^{N_{f}-1}e^{m(N_{f}-s-1)}}{(N_{f}-s-1)!}\frac{d}{dm%
%}\left( e^{-m}\frac{d}{dm}\cdots \frac{d}{dm}\left( e^{-m}I(\ell+s,m)\right)
%\right)  \notag \\
%&&+\frac{(-1)^{N_{f}}e^{-m(N_{f}-s-1)}}{(N_{f}-s-1)!}\frac{d}{dm}\left( e^{m}%
%\frac{d}{dm}\cdots \frac{d}{dm}\left( e^{m}I(\ell+s,-m)\right) \right) \Bigg) 
%\notag
%\end{eqnarray}%
These expressions help us to alternatively express \eqref{fifj_2} as a sum
of derivatives of \eqref{imor} with respect to $m$. Particularly, 
\begin{eqnarray}
\int_{-\infty }^{\infty }d\mu \frac{e^{(\ell +N_{f})\mu }e^{-\mu ^{2}/2g}}{%
(1+e^{\mu +m})^{N_{f}-s}} &=&\frac{(-1)^{N_{f}-s}e^{-(N_{f}-s)m}}{%
(N_{f}-s-1)!}\sum_{n=0}^{N_{f}-s-1}C_{N_{f}-s-1,n}^{(+)}I^{(n)}(\ell +s,m),
\\
\int_{-\infty }^{\infty }d\mu \frac{e^{(\ell +N_{f})\mu }e^{-\mu ^{2}/2g}}{%
(1+e^{\mu -m})^{N_{f}-s}} &=&\frac{e^{(N_{f}-s)m}}{(N_{f}-s-1)!}%
\sum_{n=0}^{N_{f}-s-1}C_{N_{f}-s-1,n}^{(-)}I^{(n)}(\ell +s,-m),
\end{eqnarray}%
where $I^{(n)}(\ell ,m)\equiv \frac{d^{n}I(\ell ,m)}{dm^{n}}$. The constant
factors $C_{p,q}^{(\pm )}$ are given by the recursion relations 
\begin{eqnarray}
C_{p,q}^{(+)} &=&%
\begin{cases}
-p\ C_{p-1,q}^{(+)}+C_{p-1,q-1}^{(+)}, & p>q>0, \\ 
(-1)^{p+1}p!, & p>q=0, \\ 
-1, & p=q.%
\end{cases}
\\
C_{p,q}^{(-)} &=&%
\begin{cases}
pC_{p-1,q}^{(-)}+C_{p-1,q-1}^{(-)}, & p>q>0, \\ 
p!, & p>q=0, \\ 
1, & p=q,%
\end{cases}%
\end{eqnarray}%
%
%
%
%
%
%
%
%
%
%
%
%for $p>q>0$ and $C_{i,0}^{(+)}=(-1)^{i+1}i!$, $C_{i,i}^{(+)}=-1$, $%
%C_{i,0}^{(-)}=i!$, $C_{i,i}^{(-)}=1$
respectively. To summarize 
\begin{eqnarray}
(f_{i},f_{j}) &=&\frac{c^{i+j+1}e^{-(\ln c)^{2}/2g}(-1)^{N_{f}}}{%
2^{N_{f}}(\sinh m)^{N_{f}}}\sum_{s=0}^{N_{f}-1}\binom{N_{f}+s-1}{s}\frac{1}{%
(2\sinh m)^{s}(N_{f}-s-1)!}  \notag \\
&&\times \Bigg(\sum_{n=0}^{N_{f}-s-1}C_{N_{f}-s-1,n}^{(+)}I^{(n)}(\ell +s,m)
\notag \\
&&+\sum_{n=0}^{N_{f}-s-1}C_{N_{f}-s-1,n}^{(-)}I^{(n)}(\ell +s,-m)\Bigg)
\end{eqnarray}%
where the derivatives of $I(\ell +s,m)$ are estimated from \eqref{mart} as
before. %which for $k>0$ becomes 
%\begin{equation}
%I(l,m)=2\pi e^{-i\pi (\ell+k/4)}e^{-m(\ell+k/2)+ikm^{2}/(4\pi )}G_{+}\left(
%k,1,-l-1+i\frac{km}{2\pi }-k/2\right)
%\end{equation}%
%with 
%\begin{equation}
%G_{+}\left( k,1,-l-1+i\frac{km}{2\pi }-k/2\right) =\frac{1}{e^{-km}-1}\left(
%i-\sqrt{\frac{i}{k}}\sum_{r=1}^{k}e^{\frac{i\pi }{k}(r-l-1-\frac{k}{2}+\frac{%
%ikm}{2\pi })^{2}}\right) ,  \label{G+}
%\end{equation}%
%and for $k<0$, it is 
%\begin{equation}
%I(l,m)=2\pi e^{i\pi (l-k/4)}e^{-m(l-k/2)+ikm^{2}/(4\pi )}G_{-}\left(
%-k,1,-l-1+i\frac{km}{2\pi }+k/2\right)
%\end{equation}%
%with%
%\begin{equation}
%G_{-}\left( -k,1,-l-1+i\frac{km}{2\pi }+k/2\right) =\frac{1}{1-e^{km}}\left(
%i+\sqrt{\frac{i}{k}}e^{km}\sum_{r=1}^{-k}e^{\frac{i\pi }{k}(r+l-\frac{k}{2}-%
%\frac{ikm}{2\pi })^{2}}\right) ,  \label{G-}
%\end{equation}

\section{Analytical expressions for the partition functions and some
interpretations}

\label{AbalyticalExpressions} We put the formalism we have just developed
into use and compute, as in \cite{Russo:2014bda}, specific instances of the
partition function of both models (\ref{Z}) and (\ref{Zm2}) for $N_{f}\geq1$%
. 
%We also show explicit cases with the Fayet-Iliopoulos parameter turned in.
In general, we restrict ourselves with presenting the cases comprising $U(1)$%
, $U(2)$ and $U(3)$, with $N_{f}=1,2,3$, $\eta = 0$ and also $\left\vert
k\right\vert =1,2,3$.

\subsection{Abelian partition functions}

The Abelian partition function is given directly by Mordell's integral. In
the case $N_{f}=2$ and $k$ arbitrary, we can use the first derivative of the
Mordell integral to obtain an explicit expression 
\begin{eqnarray}
Z_{k,N_{f}=2}^{U(1)}(\eta ) &=&\frac{e^{-m}}{2\pi }\left( I(\ell =-1+\mathrm{%
i}\eta ,m)-I^{\prime }(\ell =-1+\mathrm{i}\eta ,m)\right)   \notag \\
&=&%
\begin{cases}
-e^{-\mathrm{i}\pi (\mathrm{i}\eta +\frac{k}{4})}\ e^{-m(\mathrm{i}\eta +%
\frac{k}{2})+\frac{\mathrm{i}km^{2}}{4\pi }}\Bigg(\left( \mathrm{i}\eta -%
\frac{\mathrm{i}km}{2\pi }+\frac{k}{2}\right) G_{+}\left( k,1,-\mathrm{i}%
\eta +\mathrm{i}\frac{km}{2\pi }-\frac{k}{2}\right) \nonumber \\ 
\qquad \qquad -G_{+}^{\prime }\left( k,1,-\mathrm{i}\eta +\mathrm{i}\frac{km%
}{2\pi }-\frac{k}{2}\right) \Bigg),\qquad k>0, \\ 
-e^{\mathrm{i}\pi (\mathrm{i}\eta -\frac{k}{4})}\ e^{-m(\mathrm{i}\eta -%
\frac{k}{2})+\frac{\mathrm{i}km^{2}}{4\pi }}\Bigg(\left( \mathrm{i}\eta -%
\frac{\mathrm{i}km}{2\pi }-\frac{k}{2}\right) G_{-}\left( -k,1,-\mathrm{i}%
\eta +\mathrm{i}\frac{km}{2\pi }+\frac{k}{2}\right) \nonumber \\ 
\qquad \qquad -G_{-}^{\prime }\left( -k,1,-\mathrm{i}\eta +\mathrm{i}\frac{km%
}{2\pi }+\frac{k}{2}\right) \Bigg),\qquad k<0,%
\end{cases}
\label{Abelian-twoflav}
\end{eqnarray}%
where the Gauss sums are again (\ref{mert}) and (\ref{mert2}). As pointed
out in \cite{Russo:2014bda}, the formulae contain perturbative as well as
non-perturbative terms. The perturbative terms arise from the weak-coupling
expansion of factors $e^{\frac{\mathrm{i}\pi }{k}(r-1)^{2}}=e^{\frac{g}{2}%
(r-1)^{2}}$, whereas non-perturbative terms are factors $e^{\frac{\mathrm{i}%
k(m-\mathrm{i}\pi )^{2}}{4\pi }}=e^{-\frac{(m-\mathrm{i}\pi )^{2}}{2g}}$ and 
$e^{km}=e^{\frac{2\pi \mathrm{i}m}{g}}$. For specific values of $k$, and
flavour $N_{f}=1$, we obtain%
\begin{eqnarray*}
Z_{k=1,N_{f}=1}^{U(1)} &=&\frac{e^{m/2}e^{\mathrm{i}\frac{\pi }{2}}\left(
1-e^{\mathrm{i}\frac{\pi }{4}}e^{\frac{\mathrm{i}m^{2}}{4\pi }}\right) }{%
e^{m}+1}, \\
Z_{k=-1,N_{f}=1}^{U(1)} &=&\frac{e^{m/2}\left( -e^{\mathrm{i}\frac{\pi }{2}%
}+e^{\mathrm{i}\frac{\pi }{4}}e^{-\frac{\mathrm{i}m^{2}}{4\pi }}\right) }{%
e^{m}+1}, \\
Z_{k=2,N_{f}=1}^{U(1)} &=&\frac{e^{\frac{\pi \mathrm{i}}{2}}e^{m/2}\left( -%
\sqrt{2}e^{\frac{m(\mathrm{i}m+\pi )}{2\pi }}+e^{-\frac{\pi \mathrm{i}}{8}%
}\left( 1+e^{m}\right) \right) }{\sqrt{2}\left( 1+e^{2m}\right) }, \\
Z_{k=-2,N_{f}=1}^{U(1)} &=&-\frac{\sqrt{2}e^{m/2}e^{\frac{5\mathrm{i}\pi }{8}%
}(1+e^{m})-2\mathrm{i}e^{m}e^{-\frac{\mathrm{i}m^{2}}{2\pi }}}{2\left(
e^{2m}+1\right) }, \\
Z_{k=3,N_{f}=1}^{U(1)} &=&\frac{e^{3m/2}\left( 1/\sqrt{3}-e^{\mathrm{i}\pi
/4}e^{\frac{3\mathrm{i}m^{2}}{4\pi }}+\left( e^{\mathrm{i}\pi /2}+1/\sqrt{3}%
\right) \cosh m\right) }{\left( e^{3m}+1\right) }.
\end{eqnarray*}%
Notice that there seem to be apparent poles at $m=\mathrm{i}\pi /k$ and,
even though masses are real, one can still look for poles or zeros of the
partition function on the complex plane, see \cite%
{Anderson:2015ioa,Russo:2015exa} for example. As we shall see below with
more cases, these supposed poles seem to appear for $m=2\mathrm{i}\pi /k$
for $N_{f}$ \ even and $m=\mathrm{i}\pi /k$ for odd $N_{f}$. That is
actually not the case, and the partition function is smooth at these points,
as expected. We explicitly compute and check that the 
%It is immediate to check by explicit computation that also the
derivatives are smooth functions of the mass too.

As a matter of fact, it is known that the Mordell integral (\ref{imor}) is
an holomorphic function\footnote{%
We write (\ref{imor}) in the equivalent form $h(z,\tau )=\int_{%
%TCIMACRO{\U{211d} }%
%BeginExpansion
\mathbb{R}
%EndExpansion
}dx\exp (\pi \mathrm{i}\tau x^{2}-2\pi zx)/\cosh \pi x,$ then, with $z\in 
%TCIMACRO{\U{2102} }%
%BeginExpansion
\mathbb{C}
%EndExpansion
$ and $\tau \in \mathbb{H}$. It is the only holomorphic function in $z$
which satisfies \cite{Mock} $h(z)+h(z+1)=\frac{2}{\sqrt{-\mathrm{i}\tau }}%
e^{\pi \mathrm{i}\left( z+1/2\right) ^{2}/\tau }$ and $h(z)+e^{-2\pi \mathrm{%
i}z-\pi \mathrm{i}\tau }h(z+\tau )=2e^{-\pi \mathrm{i}z-\pi \mathrm{i}\tau
/4}$. It also satisfies an S-modular property, which can be used to obtain
the Giveon-Kutasov duality in the Abelian case (\ref{AGK}). Details will
appear elsewhere.} in $l$. Since the Abelian partition functions are
directly Mordell integrals, then they are holomorphic in that parameter,
which comprises both the mass\footnote{%
Although $l$ does not explicitly depend on $m$, we can promote the mass term
into the linear part of the exponential in the numerator, by shifting the
eigenvalues by the mass $m$.} and the FI parameter. The same implication
holds for the non-Abelian case, since it is a determinant of holomorphic
functions.

In the $k=1$ case, the duality is between Abelian theories and we indeed
check that%
\begin{equation}
\frac{Z_{k=1,N_{f}=1}^{U(1)}}{Z_{k=-1,N_{f}=1}^{U(1)}}=e^{\frac{\mathrm{i}%
\pi }{4}+\frac{\mathrm{i}m^{2}}{4\pi }}.  \label{AGK}
\end{equation}%
Likewise, the cases $k=2$ and $k=3$ above, will be related with duals below,
with gauge group $U(2)$ and $U(3)$ respectively. For $N_{f}=2$ we have%
\begin{eqnarray*}
Z_{k=1,N_{f}=2}^{U(1)} &=&\frac{e^{\mathrm{i}\frac{\pi }{4}}\left( -\pi +e^{%
\frac{\mathrm{i}m^{2}}{4\pi }}\left( \pi \cosh \left( m/2\right) -\mathrm{i}%
m\sinh (m/2\right) )\right) }{2\pi \left( \cosh m-1\right) }, \\
Z_{k=-1,N_{f}=2}^{U(1)} &=&\frac{-e^{m/2}e^{-\frac{\mathrm{i}m^{2}}{4\pi }%
}\left( 2\pi e^{\frac{\mathrm{i}\left( m-\mathrm{i}\pi \right) ^{2}}{4\pi }%
}+e^{\mathrm{i}\pi /4}\left( m-e^{m}m+\mathrm{i}\pi \left( 1+e^{m}\right)
\right) \right) }{2\pi \left( e^{m}-1\right) ^{2}}, \\
Z_{k=2,N_{f}=2}^{U(1)} &=&\frac{e^{\frac{m}{2}(6+\frac{\mathrm{i}m}{\pi }%
)}(\pi -\mathrm{i}m)+e^{m+\frac{\mathrm{i}m^{2}}{2\pi }}(\pi +\mathrm{i}%
m)-\left( 1-\mathrm{i}\right) \pi e^{2m}\left( \cosh m+\mathrm{i}\right) }{%
\pi \left( e^{2m}-1\right) ^{2}}, \\
Z_{k=-2,N_{f}=2}^{U(1)} &=&\frac{-\sqrt{2}e^{\frac{\pi \mathrm{i}}{4}}\pi
e^{2m}\left( \cosh m-\mathrm{i}\right) +2e^{\frac{m}{2}(4-\frac{\mathrm{i}m}{%
\pi })}(\pi \cosh m+\mathrm{i}m\sinh m)}{\pi \left( e^{2m}-1\right) ^{2}}.
\end{eqnarray*}%
It is not manifest from the form in which the partition functions are given
(for example, in the last two expressions above) but the partition functions
satisfy%
\begin{equation}
Z(N_{c},N_{f},k,m,\eta )=\overline{Z}(N_{c},N_{f},-k,m,-\eta ),  \label{Zcc}
\end{equation}%
where $\overline{Z}$ denotes the complex conjugate. In the massless case it
also holds that%
\begin{equation*}
Z(N_{c},N_{f},k,m=0,\eta )=\overline{Z}(N_{c},N_{f},-k,m=0,\eta ),
\end{equation*}
because the partition function can be shown to be an even function in the FI
parameter. We have checked these properties explicitly in all the cases
analyzed, both analytically and numerically as well up to $N_{f}=12$. A
rigorous proof is immediate, but since it has more implications, it will be
given elsewhere. As an example, let us rewrite the two examples above,
showcasing their real and imaginary parts, which we respectively name $Z_{r}$
and $Z_{I}$ 
\begin{equation*}
Z_{k=\pm 2,N_{f}=2}^{U(1)}=Z_{r}\mp \mathrm{i}Z_{I},
\end{equation*}%
where%
\begin{eqnarray*}
Z_{r} &=&\frac{e^{2m}}{\left( e^{2m}-1\right) ^{2}}\left( -1-\cosh m+2\left(
\cos \left( \frac{m^{2}}{2\pi }\right) \cosh m+\frac{m}{\pi }\sin \left( 
\frac{m^{2}}{2\pi }\right) \sinh m\right) \right) , \\
Z_{I} &=&\frac{e^{2m}}{\left( e^{2m}-1\right) ^{2}}\left( 1-\cosh m-2\left(
\sin \left( \frac{m^{2}}{2\pi }\right) \cosh m-\frac{m}{\pi }\cos \left( 
\frac{m^{2}}{2\pi }\right) \sinh m\right) \right) .
\end{eqnarray*}%
It is well-known that, in general, the partition functions computed with the
localization method are complex \cite{Closset:2012vg,Closset:2012vp},
whereas unitarity demands the partition function to be real. See \cite%
{Closset:2012vg,Closset:2012vp} for a detailed discussion. In addition, we
have seen explicitly that there is a complex conjugate property when $%
k\rightarrow -k$ and below we identify a family of partition functions whose
complex conjugate is the partition function of the inverse field theory (see 
\cite{Freed:2014iua,Witten:2015aba} for the notion of inverse field theory).

\subsection{$U(2)$ gauge group}

\label{U2_gg} We begin with the $N_{f}=1$ case and consider some duals with
the Abelian examples above

\begin{eqnarray}
Z_{k=1,N_{f}=1}^{U(2)} &=&e^{-\frac{3\pi \mathrm{i}}{4}+\frac{\mathrm{i}
m^{2}}{4\pi }},  \label{marginal} \\
Z_{k=2,N_{f}=1}^{U(2)} &=&\frac{e^{-\frac{\mathrm{i}\pi }{4}}\left( 2e^{m}-%
\sqrt{2}e^{\frac{\mathrm{i}\pi }{8}}e^{\frac{\mathrm{i} m^{2}}{2\pi }%
}e^{m/2}(1+e^{m})\right) }{2\left( e^{2m}+1\right) },  \notag \\
Z_{k=-2,N_{f}=1}^{U(2)} &=&\frac{e^{\frac{\mathrm{i}\pi }{4}}e^{\frac{m}{2}-%
\frac{\mathrm{i} m^{2}}{2\pi }}\left( 2e^{\frac{m}{2}+\frac{\mathrm{i} m^{2}%
}{2\pi }}+\sqrt{2}e^{\frac{7\mathrm{i}\pi }{8}}(1+e^{m})\right) }{2\left(
e^{2m}+1\right) },  \notag \\
Z_{k=-3,N_{f}=1}^{U(2)} &=&\frac{\mathrm{i}-e^{\frac{5\mathrm{i}\pi }{12}-%
\frac{3\mathrm{i} m^{2}}{4\pi }}}{\sqrt{3}(-1+2\cosh m)}.  \notag
\end{eqnarray}%
%
%
%
%
%
%
%We obtain now the duals%
Some duality cases are 
\begin{equation*}
\frac{Z_{k=2,N_{f}=1}^{U(1)}}{Z_{k=-2,N_{f}=1}^{U(2)}}=e^{-\frac{3\pi 
\mathrm{i}}{4}+\frac{\mathrm{i} m^{2}}{2\pi }},
\end{equation*}%
and 
\begin{equation*}
\frac{Z_{k=-2,N_{f}=1}^{U(1)}}{Z_{k=2,N_{f}=1}^{U(2)}}=e^{\frac{3\pi \mathrm{%
i}}{4}-\frac{\mathrm{i} m^{2}}{2\pi }},
\end{equation*}%
as expected, because of the complex conjugation of the partition function
under the change $k\rightarrow -k.$ Let us now present some $U(2)$ examples
with $N_{f}=2$, and relate them with some of their duals above, mostly with $%
U(1)$ and $N_{f}=2$ cases%
\begin{eqnarray*}
Z_{k=1,N_{f}=2}^{U(2)} &=&\frac{e^{m/2}e^{\frac{\mathrm{i} m^{2}}{4\pi }%
}\left( m(1-e^{m})+\mathrm{i}\pi \left( 1+e^{m}-2e^{\frac{m}{2}+\frac{%
\mathrm{i} m^{2}}{4\pi }}\right) \right) }{2\pi \left( e^{m}-1\right) ^{2}},
\\
Z_{k=-1,N_{f}=2}^{U(2)} &=&\frac{e^{m/2}e^{-\frac{\mathrm{i} m^{2}}{2\pi }%
}\left( 2\pi \mathrm{i} e^{m/2}+e^{\frac{\mathrm{i} m^{2}}{4\pi }}\left(
m(1-e^{m})-\mathrm{i}\pi \left( 1+e^{m}\right) \right) \right) }{2\pi \left(
e^{m}-1\right) ^{2}}.
\end{eqnarray*}%
We can combine these particular cases to highlight a few more explicit
analytical examples of Giveon-Kutasov duality. Namely, 
\begin{eqnarray*}
\frac{Z_{k=1,N_{f}=2}^{U(1)}}{Z_{k=-1,N_{f}=2}^{U(2)}} &=& e^{\frac{3\pi 
\mathrm{i}}{4}+\frac{\mathrm{i} m^{2}}{2\pi }}, \\
\frac{Z_{k=-1,N_{f}=2}^{U(1)}}{Z_{k=1,N_{f}=2}^{U(2)}} &=& e^{-\frac{3\pi 
\mathrm{i}}{4}-\frac{\mathrm{i} m^{2}}{2\pi }}.
\end{eqnarray*}

\subsection{Supersymmetry breaking and partition functions of modulus 1}

\label{SuSy_Breaking} Notice the special form of the partition function %
\eqref{marginal} (see also \cite[Eq (2.36)]{Russo:2014bda}). Such partition
functions arise when the dual is actually a theory with $U(N_{c}=0)$ and
hence the partition function of the dual is just $1$. Therefore, the only
term remaining is the phase factor of the Giveon-Kutasov duality (\ref{GK}).

Thus, there is a family of partition functions, satisfying $\left\vert
k\right\vert -N_{c}+N_{f}=0,$ whose value are just Giveon-Kutasov phases.
This family of partition functions therefore constitute a \textit{marginal}
case, separating the partition functions which are identically zero, namely
those which satisfy $\left\vert k\right\vert -N_{c}+N_{f}<0$, due to
spontaneous supersymmetry breaking \cite{Morita:2011cs} and the regular ones
(that satisfy $\left\vert k\right\vert -N_{c}+N_{f}>0$). Our determinantal
method indeed directly gives null results for those cases characterized by $%
\left\vert k\right\vert -N_{c}+N_{f}<0$. In this way for example, we
obtained 
\begin{eqnarray*}
Z_{k=\pm 1,N_{f}=1}^{U(4)} &=&Z_{k=\pm 2,N_{f}=1}^{U(4)}=0\text{ \ and \ \ }%
Z_{k=\pm 3,N_{f}=1}^{U(4)}=e^{\pm \frac{\mathrm{i}\pi }{12}\pm \frac{3%
\mathrm{i}m^{2}}{4\pi }}, \\
Z_{k=\pm 1,N_{f}=1}^{U(5)} &=&Z_{k=\pm 2,N_{f}=1}^{U(5)}=Z_{k=\pm
3,N_{f}=1}^{U(5)}=0\text{ \ \ and \ \ }Z_{k=\pm 4,N_{f}=1}^{U(5)}=e^{\pm 
\frac{7\mathrm{i}\pi }{4}\pm \frac{\mathrm{i}m^{2}}{\pi }}
\end{eqnarray*}%
and so on. 
%In particular, we also show below, when we present examples for the $U(3)$ case with $N_{f}=1,$ that we consecutively obtain a null result, a pure GK phase and a normal partition function, for the cases $k=1,2$ and $3$, as expected. 
It is noteworthy that this type of partition function, being a complex
number of modulus one, is the one that emerges in the description of
symmetry protected phases \cite{Chen:2011pg}. These partition functions $Z_{%
\mathbb{S}^{3}}=e^{\mathrm{i}\Phi }$ are %partition functions 
of a topological quantum field theory which is $\mathit{invertible}$ \cite%
{Freed:2014iua, Witten:2015aba}, with its inverse being the theory with
complex conjugate partition function $Z_{\mathbb{S}^{3}}^{-1}=e^{-\mathrm{i}%
\Phi }$, which in our case corresponds to $k\rightarrow -k$. Thus, for our $%
\mathcal{N}=2$ theory, through the Giveon-Kutasov duality, we have seen that
the partition functions that exhibit such behavior come exclusively from the
anomaly phase factor. This is also consistent with the idea that anomalies
are invertible field theories (\cite{Freed:2014iua} and references therein).
Further analysis of this result here from the perspective of study of the
topological phases of matter seems an interesting open problem. In the next
Section, we actually give an analytical expression for $\Phi =\Phi
(k,N_{c},N_{f},m,\eta )$.

\subsection{$U(3)$ gauge group}

For $N_{f}=1$ we have%
\begin{eqnarray*}
Z_{k=1,N_{f}=1}^{U(3)} &=& 0, \\
Z_{k=2,N_{f}=1}^{U(3)} &=& e^{-\frac{3\pi \mathrm{i}}{4}+\frac{\mathrm{i}
m^{2}}{2\pi }}, \\
Z_{k=3,N_{f}=1}^{U(3)} &=& \frac{e^{\frac{3\pi \mathrm{i}}{4}} e^{m/2}
\left(6e^{-\frac{11\pi \mathrm{i}}{12}} e^m+2 \sqrt{3} e^{\frac{m}{4}
\left(8+\frac{3 \mathrm{i} m}{\pi }\right)}+\left(3 \mathrm{i}+\sqrt{3}%
\right) e^{m+\frac{3 \mathrm{i} m^2}{4 \pi }}+2 \sqrt{3} e^{\frac{3 \mathrm{i%
} m^2}{4 \pi }}\right)}{6 \left(e^{3 m}+1\right)} , \\
Z_{k=-3,N_{f}=1}^{U(3)} &=& \frac{\sqrt{2}e^{-\frac{3\pi \mathrm{i}}{4}} e^{%
\frac{3}{4} m \left(4-\frac{\mathrm{i} m}{\pi }\right)} \cosh(m/2) (2 \cosh
m-1)}{6\left(e^{3 m}+1\right)^2} \Big(2\sqrt{6} e^{-\frac{\pi \mathrm{i}}{3}%
}-6\sqrt{2}e^{-\frac{\pi \mathrm{i}}{12}}e^{\frac{3 \mathrm{i} m^2}{4 \pi }}+
\notag \\
& & 4 \sqrt{6} \cosh m\Big).
\end{eqnarray*}%
%
%
%
%
%
%
%
%
%The duals are 
For these values of the parameters the duality now becomes 
\begin{eqnarray*}
\frac{Z_{k=2,N_{f}=1}^{U(3)}}{ Z_{k=2,N_{f}=1}^{U(0)}} &=& e^{-\frac{3\pi 
\mathrm{i}}{4}+\frac{\mathrm{i} m^{2}}{2\pi }}, \\
\frac{Z_{k=3,N_{f}=1}^{U(1)}}{Z_{k=-3,N_{f}=1}^{U(3)}}&=& e^{-\frac{11\pi 
\mathrm{i}}{12}+\frac{3\mathrm{i} m^{2}}{4\pi }}.
\end{eqnarray*}%
For $N_{f}=2$ we have%
\begin{eqnarray*}
Z_{k=1,N_{f}=2}^{U(3)} &=& e^{\frac{3\pi \mathrm{i}}{4}+\frac{\mathrm{i}
m^{2}}{2\pi }}, \\
Z_{k=2,N_{f}=2}^{U(3)} &=& \frac{e^{m+\frac{\mathrm{i} m^2}{2 \pi }} }{\sqrt{%
2}\pi \left(e^{2 m}-1\right)^2 }\Big(\sqrt{2}\left(1-e^{2 m}\right) m+\sqrt{2%
}\pi e^{\frac{\pi \mathrm{i}}{2}}\left(1+ e^{2 m}\right) +  \notag \\
& & \pi e^{-\frac{\pi \mathrm{i}}{4}} e^{\frac{1}{2} m \left(4+\frac{\mathrm{%
i} m}{\pi }\right)}-2 \pi e^{\frac{\pi \mathrm{i}}{4}}e^{m+\frac{\mathrm{i}
m^2}{2 \pi }} + \pi e^{-\frac{\pi \mathrm{i}}{4}} e^{\frac{\mathrm{i} m^2}{2
\pi }}\Big) , \\
Z_{k=3,N_{f}=2}^{U(3)} &=&\frac{e^{\frac{3\mathrm{i} m^{2}}{4\pi }}}{\sinh
\left( \frac{3m}{2}\right) }\left( \frac{\sqrt{3}m}{4\pi }e^{\frac{7\pi 
\mathrm{i}}{12}}+\frac{e^{\frac{\pi \mathrm{i}}{2}}}{2\sqrt{2}}\coth \left( 
\frac{3m}{2}\right) \right) +\frac{e^{7\pi \mathrm{i}/12}}{4\sinh ^{2}\left( 
\frac{3m}{2}\right) } \\
& &+\frac{e^{\frac{3\mathrm{i} m^{2}}{4\pi }}}{\sqrt{3}\sinh ^{2}\left( 
\frac{3m}{2}\right) }\left( e^{7\pi \mathrm{i}/4}\left( \frac{1}{2}+\cosh
\left( \frac{m}{2}\right) \right) +\frac{\cosh \left( m\right) }{2}\right) .
\end{eqnarray*}%
The duals we extract from these cases are%
\begin{equation*}
\frac{Z_{k=-2,N_{f}=2}^{U(1)}}{Z_{k=2,N_{f}=2}^{U(3)}}=e^{\frac{3\pi \mathrm{%
i}}{2}-\frac{\mathrm{i} m^{2}}{\pi }} \qquad \text{and}\qquad \frac{%
Z_{k=2,N_{f}=2}^{U(1)}}{Z_{k=-2,N_{f}=2}^{U(3)}}=e^{\frac{\pi \mathrm{i}}{2}+%
\frac{\mathrm{i} m^{2}}{\pi }},
\end{equation*}%
and we also find%
\begin{equation*}
\frac{Z_{k=3,N_{f}=2}^{U(2)}}{Z_{k=-3,N_{f}=2}^{U(3)}}=e^{-\frac{11\pi 
\mathrm{i}}{12}+\frac{3\mathrm{i} m^{2}}{2\pi }}.
\end{equation*}

\subsection{Cases with $N_{f}=3$}

Instead of giving the explicit partition functions for the $N_{f}=3$ case,
we give the ratio of the dual partition functions directly%
\begin{eqnarray*}
\frac{Z_{k=1,N_{f}=3}^{U(2)}}{Z_{k=-1,N_{f}=3}^{U(2)}} &=& e^{\frac{3\mathrm{%
i} \pi }{4}+\frac{3\mathrm{i} m^{2}}{4\pi }}, \\
\frac{Z_{k=2,N_{f}=3}^{U(2)}}{Z_{k=-2,N_{f}=3}^{U(3)}} &=& e^{\frac{\mathrm{i%
}\pi }{4}+\frac{3\mathrm{i} m^2}{2\pi}}.
\end{eqnarray*}%
Note that the duals in the former case have the same gauge group. It is
immediate to check that this is always the case when, for $m,n\in 
%TCIMACRO{\U{2115} }%
%BeginExpansion
\mathbb{N}
%EndExpansion
$ (or, more generally $n\in 
%TCIMACRO{\U{2124} }%
%BeginExpansion
\mathbb{Z}
%EndExpansion
$ if $\left\vert n\right\vert <m$) we have%
\begin{equation}
N_{c}=m+n,\text{ \ }k=m\text{ \ and \ }N_{f}=m+2n,  \label{set}
\end{equation}%
because then $Z(m+n,m+2n,m)=e^{i\pi \phi _{\mathrm{GK}}}Z(m+n,m+2n,-m).$ In
addition, since the partition function on the r.h.s. is the complex
conjugate of the one in the l.h.s., if we write $Z(m+n,m+2n,m)=re^{\mathrm{i}%
\theta }$ in polar form, we see that for partition functions characterized
by (\ref{set}) it holds that $\theta =\phi _{\mathrm{GK}}/2$.

\section{Giveon-Kutasov duality}

The Giveon-Kutasov duality is between $U(N_{c})$ and $U(|k|+N_{f}-N_{c})$,
where $k$ is the Chern-Simons level. In particular, the theories are:

\begin{itemize}
\item $\mathcal{N}=2$ $U(N_c)$ gauge theory with $N_f$ flavors and a
Chern-Simons term at level $k$.

\item $\mathcal{N}=2$ $U(|k|+N_{f}-N_{c})$ gauge theory with $N_{f}$ flavors
and a Chern-Simons term at level $-k$. In addition, there are ${N_{f}^{2}}$
uncharged chiral multiplets ${M_{a}}^{b}$, which couple through a
superpotential $\tilde{q}^{a}{M_{a}}^{b}q_{b}$.
\end{itemize}

As explained in the Introduction, the Giveon-Kutasov duality specifically
implies for the partition function%
\begin{equation}
Z_{N_{f},k}^{U(N_{c})}\left( \eta \right) =e^{\mathrm{sgn}(k)\pi \mathrm{i}
\left( c_{\left\vert k\right\vert ,N_{f}}-\eta ^{2}\right)
}Z_{N_{f},-k}^{U(\left\vert k\right\vert +N_{f}-N_{c})}\left( -\eta \right) ,
\label{Zduality}
\end{equation}%
where $c\left( \left\vert k\right\vert ,N_{c},N_{f}\right) $ is a polynomial
quadratic in the level $k$ (or rather on its absolute value) and the
coefficients have a non-trivial dependence on $N_{f}$ (and, we find, on $%
N_{c}$ as well). As discussed in \cite{Closset:2012vp}, this phase can be
attributed to certain contact terms that must be added to the action to
ensure reflection positivity. Thus, using the matrix model representation of
the partition function of the $\mathcal{N}=3$, $U(N_{c})$ theory with $N_{f}$
fundamental flavors\footnote{%
Setting $\mu _{i}=2\pi \lambda _{i}$ and $m_{a}=m$ for all $a=1,2,\ldots
,N_{f}$ one obtains \eqref{Z}.} \cite{Kapustin:2013hpk}%
\begin{equation}
Z_{k,N_{f},N_{c}}(\eta ,m_{a})=e^{\mathrm{i}\delta (N_{c},k,N_{f};\eta
,m_{a})}\frac{1}{N_{c}!}\int d^{N_{c}}\lambda \prod_{j=1}^{N_{c}}\frac{%
e^{-k\pi \mathrm{i}{\lambda _{j}}^{2}+2\pi \mathrm{i}\eta \lambda _{j}}}{%
\prod_{a=1}^{N_{f}}2\cosh \pi (\lambda _{j}+m_{a})}\prod_{i<j}\left( 2\sinh
\pi (\lambda _{i}-\lambda _{j})\right) ^{2},  \label{m}
\end{equation}%
where, following the notation and presentation in \cite{Kapustin:2013hpk}, $%
\delta $ is chosen so that $Z$ is real and positive, then the statement (\ref%
{Zduality}) is encapsulated in%
\begin{equation}
Z_{k,N_{f},N_{c}}(\eta ,m_{a})=Z_{-k,N_{f},|k|+N_{f}-N_{c}}(-\eta ,m_{a}).
\label{kw-d}
\end{equation}%
In \cite{Kapustin:2013hpk}, it is argued that an explicit formula for the
relative phase can be computed by studying the contact terms of the dual
theories \cite{Closset:2012vp} and the expression%
\begin{align}
\gamma (N_{c},k,N_{f};\eta ,m_{a})& :=\delta (|k|+N_{f}-N_{c},-k,N_{f};-\eta
,m_{a})-\delta (N_{c},k,N_{f};\eta ,m_{a})  \notag \\
& =\frac{1}{24}(k^{2}+3(k+N_{f})(N_{f}-2)+2)+\frac{1}{2}\eta ^{2}-\frac{1}{2}%
k\sum_{a}{m_{a}}^{2}-\eta \sum_{a}m_{a}  \label{m2}
\end{align}%
is provided. Its derivation appeared in the previous work \cite%
{Willett:2011gp}. The formulas in \cite{Willett:2011gp} and \cite%
{Kapustin:2013hpk} are very similar with the exception of a global $-2\pi $
multiplicative re-scaling, which indeed seems to be missing in (\ref{m})-(%
\ref{m2}). Additionally, the expression in \cite{Willett:2011gp} does not
contain the last term in \eqref{m2}, involving both the FI parameter and the
masses\footnote{%
The duality considered in \cite{Willett:2011gp} is between $%
Z_{k,N_{f},N_{c}}(\eta ,m)$ and $Z_{-k,N_{f},|k|+N_{f}-N_{c}}(\eta ,-m)$
instead of eq. \eqref{kw-d}.}. %(at least certainly the $\pi $). 
Former work of the same authors conjectured that \cite{Kapustin:2010mh}%
\begin{equation}
Z_{k,N_{f}}^{(N_{c})}(\eta ;m_{a})=e^{\text{sgn}(k)\pi \mathrm{i}%
(c_{|k|,N_{f}}-\eta ^{2})}e^{\sum_{a}(k\pi \mathrm{i}{m_{a}}^{2}+2\pi 
\mathrm{i}\eta m_{a})}Z_{-k,N_{f}}^{(|k|+N_{f}-N_{c})}(-\eta ;m_{a}),
\label{conj2}
\end{equation}%
where%
\begin{equation}
c_{k,N_{f}}=-\frac{1}{12}(k^{2}+3(N_{f}-2)k+a_{N_{f}})  \label{QF2}
\end{equation}%
with%
\begin{equation*}
a_{N_{f}}=\left\{ 
\begin{array}{lll}
-1, & N_{f}=1,\text{ }(\mbox{mod }4), &  \\ 
2, & N_{f}=2,4\text{ }(\mbox{mod }4), &  \\ 
-13, & N_{f}=3\text{ }(\mbox{mod }4). & 
\end{array}%
\right.
\end{equation*}%
It is mentioned in \cite{Willett:2011gp} that there is consistency between
the two formulas, but in general they do give different results for the
phase factor. For example, while the two formulas agree for $N_{f}=1$ and $%
N_{f}=2$ and generic $k$, they differ for $N_{f}=3$. For instance%
\begin{equation*}
-2\pi \gamma (N_{c},1,3,\eta =0,m_{a}=0)=-\frac{5}{4}\pi \text{ \ whereas \
\ }\pi c_{1,3}=\frac{3}{4}\pi .
\end{equation*}%
Note also that the factor sgn$(k)$ that appears in (\ref{conj2}) but not in
the other two, above mentioned, expressions, guarantees consistency if one
applies the duality again on the r.h.s. of (\ref{conj2}).

We %Thus, it is specially convenient that we can 
further test the duality, in the $\mathcal{N}=3$ setting, 
%by using our exact analysis 
of the matrix model \eqref{m} using the formalism developed in Section \ref%
{NfGeneric}.

\subsection{Explicit expression for the phase factor}

\begin{table}[tbp]
\begin{tabular}{|c|c|c|c|c|c|c|c|c|c|c|c|}
\hline
&  & \multicolumn{10}{|c|}{$k$} \\ \hline
$N_f$ & $N_c$ & -5 & -4 & -3 & -2 & -1 & 1 & 2 & 3 & 4 & 5 \\ \hline\hline
1 & 1 & 3/4 & 1/4 & 11/12 & 3/4 & -1/4 & 1/4 & -3/4 & -11/12 & -1/4 & -3/4
\\ \hline
1 & 2 & -1/4 & 1/4 & -1/12 & 3/4 & 3/4 & -3/4 & -3/4 & 1/12 & -1/4 & 1/4 \\ 
\hline
1 & 3 & -3/4 & 1/4 & 11/12 & 3/4 & NaN & NaN & -3/4 & -11/12 & -1/4 & -3/4
\\ \hline
2 & 1 & -3/4 & -1/2 & 11/12 & -1/2 & -3/4 & 3/4 & 1/2 & -11/12 & 1/2 & 3/4
\\ \hline
2 & 2 & -3/4 & 1/2 & 11/12 & 1/2 & -3/4 & 3/4 & -1/2 & -11/12 & -1/2 & 3/4
\\ \hline
2 & 3 & -3/4 & -1/2 & 11/12 & -1/2 & -3/4 & 3/4 & 1/2 & -11/12 & 1/2 & 3/4
\\ \hline
2 & 4 & -3/4 & 1/2 & 11/12 & 1/2 & NaN & NaN & -1/2 & -11/12 & -1/2 & 3/4 \\ 
\hline
3 & 1 & -3/4 & 1/4 & 5/12 & -1/4 & 1/4 & -1/4 & 1/4 & -5/12 & -1/4 & 3/4 \\ 
\hline
3 & 2 & 1/4 & 1/4 & -7/12 & -1/4 & -3/4 & 3/4 & 1/4 & 7/12 & -1/4 & -1/4 \\ 
\hline
3 & 3 & -3/4 & 1/4 & 5/12 & -1/4 & 1/4 & -1/4 & 1/4 & -5/12 & -1/4 & 3/4 \\ 
\hline
3 & 4 & 1/4 & 1/4 & -7/12 & -1/4 & -3/4 & 3/4 & 1/4 & 7/12 & -1/4 & -1/4 \\ 
\hline
3 & 5 & -3/4 & 1/4 & 5/12 & -1/4 & NaN & NaN & 1/4 & -5/12 & -1/4 & 3/4 \\ 
\hline
4 & 1 & 3/4 & 1/2 & -7/12 & -1/2 & 3/4 & -3/4 & 1/2 & 7/12 & -1/2 & -3/4 \\ 
\hline
4 & 2 & 3/4 & -1/2 & -7/12 & 1/2 & 3/4 & -3/4 & -1/2 & 7/12 & 1/2 & -3/4 \\ 
\hline
4 & 3 & 3/4 & 1/2 & -7/12 & -1/2 & 3/4 & -3/4 & 1/2 & 7/12 & -1/2 & -3/4 \\ 
\hline
4 & 4 & 3/4 & -1/2 & -7/12 & 1/2 & 3/4 & -3/4 & -1/2 & 7/12 & 1/2 & -3/4 \\ 
\hline
4 & 5 & 3/4 & 1/2 & -7/12 & -1/2 & 3/4 & -3/4 & 1/2 & 7/12 & -1/2 & -3/4 \\ 
\hline
5 & 1 & -1/4 & 1/4 & -1/12 & 3/4 & 3/4 & -3/4 & -3/4 & 1/12 & -1/4 & 1/4 \\ 
\hline
5 & 2 & 3/4 & 1/4 & 11/12 & 3/4 & -1/4 & 1/4 & -3/4 & -11/12 & -1/4 & -3/4
\\ \hline
5 & 3 & -1/4 & 1/4 & -1/12 & 3/4 & 3/4 & -3/4 & -3/4 & 1/12 & -1/4 & 1/4 \\ 
\hline
5 & 4 & 3/4 & 1/4 & 11/12 & 3/4 & -1/4 & 1/4 & -3/4 & -11/12 & -1/4 & -3/4
\\ \hline
5 & 5 & -1/4 & 1/4 & -1/12 & 3/4 & 3/4 & -3/4 & -3/4 & 1/12 & -1/4 & 1/4 \\ 
\hline
6 & 1 & 1/4 & -1/2 & -1/12 & -1/2 & 1/4 & -1/4 & 1/2 & 1/12 & 1/2 & -1/4 \\ 
\hline
6 & 2 & 1/4 & 1/2 & -1/12 & 1/2 & 1/4 & -1/4 & -1/2 & 1/12 & -1/2 & -1/4 \\ 
\hline
6 & 3 & 1/4 & -1/2 & -1/12 & -1/2 & 1/4 & -1/4 & 1/2 & 1/12 & 1/2 & -1/4 \\ 
\hline
6 & 4 & 1/4 & 1/2 & -1/12 & 1/2 & 1/4 & -1/4 & -1/2 & 1/12 & -1/2 & -1/4 \\ 
\hline
6 & 5 & 1/4 & -1/2 & -1/12 & -1/2 & 1/4 & -1/4 & 1/2 & 1/12 & 1/2 & -1/4 \\ 
\hline
7 & 1 & 1/4 & 1/4 & -7/12 & -1/4 & -3/4 & 3/4 & 1/4 & 7/12 & -1/4 & -1/4 \\ 
\hline
8 & 1 & -1/4 & 1/2 & 5/12 & -1/2 & -1/4 & 1/4 & 1/2 & -5/12 & -1/2 & 1/4 \\ 
\hline
9 & 1 & 3/4 & 1/4 & 11/12 & 3/4 & -1/4 & 1/4 & -3/4 & -11/12 & -1/4 & -3/4
\\ \hline
11 & 1 & -3/4 & 1/4 & 5/12 & -1/4 & 1/4 & -1/4 & 1/4 & -5/12 & -1/4 & 3/4 \\ 
\hline
\end{tabular}
\vspace{3mm}
\caption{Values of the parameter $\protect\theta$ for different values of $%
N_f, N_c$ and $k$ for the theory with $N_f$ massless hypermultiplets, i.e.
matrix model \eqref{Z}. NaN refers to instances where the dual theory is not
well defined because the dual number of colours is negative, i.e. $|k| + N_f
- N_c < 0$, see discussion in Section \protect\ref{SuSy_Breaking}.}
\label{tbl1}
\end{table}

For $N_{f}>1$, it becomes computational intractable to estimate \eqref{Z} in
pen and paper. Thus, we programmed \eqref{Z}, using \textit{Mathematica}, as
a function which takes as input parameters the variables $%
(k,N_{f},N_{c},m,\eta )$. 
%Since we have computed (\ref{m}) exactly for a multitude of cases we can test these formulas for higher $N_{f}$.  
These computations are symbolical and work well for low values of $%
N_{f},N_{c},k$, however for larger values, the symbolical calculations
become time consuming \footnote{%
All computations were performed on a laptop with Intel Core2 Duo CPU T6400 
%TCIMACRO{\TeXButton{@}{\@}}%
%BeginExpansion
\@%
%EndExpansion
2.00 GHz and 3GB RAM and a Fujitsu Server Primergy TX100 S3 with 8
processors and 8GB RAM.}. As a first step, we solved symbolically \eqref{Z}
for low values $N_{f},N_{c}$, see also Section \ref{AbalyticalExpressions}.
We verified that neither the mass term nor the FI term couple with $k$, as
expected and suggested in \eqref{m2} and \eqref{conj2}. Therefore, to
further investigate the form of the quadratic function in $k$ we focus on
the massless and $\eta =0$ case. This is convenient because floating-point
arithmetic and parallelization methods on \textit{Mathematica} scripts,
speed up our code and enable us to compute cases up to $N_{f}=12$.

Typically, our process is the following: for specific values of $N_{f}$ and $%
N_{c}$ we find the ratios 
\begin{equation}
\frac{Z_{N_{f},k}^{U(N_{c})}\left( 0\right) }{Z_{N_{f},-k}^{U(\left\vert
k\right\vert +1N_{f}-N_{c})}\left( 0\right) }=e^{\mathrm{i}\pi \theta }
\end{equation}%
for $0<|k|\leq 5$. The ratio is always a phase, e.g. $e^{\mathrm{i}\pi
\theta },$ $\theta \in \mathbb{R}$. We then determine a quadratic function
in $|k|$, $\phi _{N_{f},N_{c}}(k)$, that captures all phases for $0<|k|\leq
5 $. We repeat for different values of $N_{f}$ and $N_{c}$ and certain
patterns for the quadratic functions $\phi _{N_{f},N_{c}}(|k|)$ arise.

We summarize our results in Table \ref{tbl1}, where we present the parameter 
$\theta $ for different values of $N_{f},N_{c}$ and $0<|k|\leq 5$. We
observe the anti-symmetry between negative and positive $k$, which leads us
to write $e^{i\pi \theta }=e^{\text{sgn}(k)\mathrm{i}\pi \phi
_{N_{f},N_{c}}(|k|)}$. This confirms the phase in \eqref{Zduality} when $%
\eta =0$ and $m=0$. First, we see that neither \eqref{m2} nor \eqref{QF2}
are good candidates for reproducing the values in Table \ref{tbl1}. In
particular, applying \eqref{m2} and \eqref{conj2} with \eqref{QF2} for $%
N_{f}=1$ and $k=1,2,\ldots ,5$ one finds 
\begin{equation*}
\theta =%
\begin{cases}
\frac{1}{4}, & k=1, \\ 
\frac{1}{4}, & k=2, \\ 
\frac{1}{12}, & k=3, \\ 
-\frac{1}{4}, & k=4, \\ 
-\frac{3}{4}, & k=5,%
\end{cases}%
\end{equation*}%
respectively. These values are different from those presented in Table \ref%
{tbl1}. Therefore we need to go beyond the existing conjectured quadratic
functions and find a new one. Thus, we search for a universal quadratic
function 
\begin{equation}
\phi (N_{f},N_{c},|k|)=\alpha (N_{f},N_{c})k^{2}+\beta
(N_{f},N_{c})|k|+\gamma (N_{f},N_{c}),  \label{universalquadratic}
\end{equation}%
which captures all the values obtained. This is done in two steps. In the
first step, for each $N_{f}$ and $N_{c}$ we use the $\theta $ values for $%
k=1,2,3$ to find a quadratic function. That is, we solve the system of
equations 
\begin{eqnarray}
\theta _{1} &=&\alpha _{N_{f},N_{c}}+\beta _{N_{f},N_{c}}+\gamma
_{N_{f},N_{c}}  \notag \\
\theta _{2} &=&4\alpha _{N_{f},N_{c}}+2\beta _{N_{f},N_{c}}+\gamma
_{N_{f},N_{c}}  \notag \\
\theta _{3} &=&9\alpha _{N_{f},N_{c}}+3\beta _{N_{f},N_{c}}+\gamma
_{N_{f},N_{c}}  \notag
\end{eqnarray}%
to find the parameters $\alpha _{N_{f},N_{c}},\beta _{N_{f},N_{c}},\gamma
_{N_{f},N_{c}}$. It is worth mentioning that there is not a unique quadratic
function that gives rise to the same phase. Had we solved for $k=2,3,4$ we
would have found different parameters which still give the same overall
phase. Therefore, we find different quadratic functions for different values
of $N_{f},N_{c}$. For example, for $N_{f}=N_{c}=1$ we find $\phi _{1,1}(k)=%
\frac{5}{12}k^{2}-\frac{9}{4}k+\frac{25}{12}$ which gives $e^{\text{sgn}%
(k)\pi \mathrm{i}\phi _{1,1}(|k|)}/e^{\pi \mathrm{i}\theta }=1$ for all $%
\theta $ in the first row of Table~\ref{tbl1}. We further do some
\textquotedblleft blind\textquotedblright\ tests computing $\theta $ for a $%
k>|5|$ and verifying the correctness of the expression $\phi
_{N_{f},N_{c}}(k)$.

In the second step, we attempt to combine these different quadratic
functions into a single quadratic function, such as %
\eqref{universalquadratic}. Our result is 
\begin{equation}
\phi \left( N_{f},N_{c},k\right) =\frac{1}{12}\left( 5k^{2}+\beta \left(
N_{f},N_{c}\right) k+\gamma \left( N_{f},N_{c}\right) \right)
\label{QuadraticFunction}
\end{equation}%
where 
\begin{eqnarray}
\beta \left( N_{f},N_{c}\right) &=&(-1)^{N_{f}}3\left( N_{f}+c+4\left(
N_{c}-1\right) \right) ,  \notag \\
\gamma \left( N_{f},N_{c}\right) &=&12\left( N_{f}-1\right) \left(
N_{c}-1\right) +%
\begin{cases}
-2, & N_{f}\mod 4=2, \\ 
10, & N_{f}\mod 4=0, \\ 
1, & N_{f}\mod 4=1\quad \text{or}\quad 3,%
\end{cases}%
\end{eqnarray}%
and 
\begin{equation*}
c=%
\begin{cases}
4, & N_{f}\!\!\mod 4=0, \\ 
0, & \text{otherwise}.%
\end{cases}%
\end{equation*}%
The quadratic function \eqref{QuadraticFunction} captures all the phases
presented in the Table~\ref{tbl1}, meaning that $e^{\text{sgn}(k)\mathrm{i}%
\pi \phi \left( N_{f},N_{c},|k|\right) }/e^{\mathrm{i}\pi \theta }=1$. This
expression is further tested as follows. As explained above, in step 1 we
have found a quadratic function for each $N_{f},N_{c}$. This function is
different from \eqref{QuadraticFunction} and different for each $N_{f},N_{c}$%
. Therefore, we further test the equality%
\begin{equation*}
e^{\text{sgn}(k)\mathrm{i}\pi \phi \left( N_{f},N_{c},|k|\right) }/e^{\text{%
sgn}(k)\mathrm{i}\pi \phi _{N_{f},N_{c}}(|k|)}=1
\end{equation*}
for $|k|>5$.

We do not claim that our result is the only valid quadratic function. It is
the simplest one we could find for which $e^{\text{sgn}(k)\mathrm{i}\pi \phi
\left( N_{f},N_{c},|k|\right) }/e^{\mathrm{i}\pi \theta }=1$. There might be
other quadratic functions $\phi (N_{f},N_{c},k)$ which reproduce our results
in Table \ref{tbl1}. As an open problem, it would be interesting to compare
such results with a full computation coming from the complete analysis of
supersymmetric Chern-Simons counterterms, since they characterize the
anomaly \cite{Closset:2012vg,Closset:2012vp}. In \cite{Benini:2011mf} (see
also \cite{Amariti:2014lla}) all the required counterterms are explicitly
given, actually for a much more general setting, including \textit{chiral}
theories, described by the matrix model (\ref{Zds}). By taking the
parameters $s_{1},s_{2}$ in \cite{Benini:2011mf} as $s_{1}=s_{2}=N_{f}$ \
one finds that the explicit expressions of the counterterms have a similar
dependence, in appearance, in $N_{c}$ and $N_{f}$ \ to the one obtained
here, although without the modular arithmetic (mod 4) behavior obtained
here. The combination of the counterterms that give the phase factor is also
well-known in general (see \cite[Eq. 5.13]{Benini:2011mf} or \cite[Eq. A.15]%
{Amariti:2014lla}), but we leave %the problem of precisely comparing the
the eventual comparison of the results obtained here with matrix models with
a direct explicit computation of the phase with the Chern-Simons
counterterms \cite{Benini:2011mf,Amariti:2014lla} as an open question for
further work.

\subsection{Massive hypemultiplets and non-zero Fayet-Iliopoulos term}

Having determined the quadratic $k$-dependence of the phase factor we
investigate the dependence on the mass, $m$, and the Fayet-Iliopoulos, $\eta 
$, terms in \eqref{Z}. As before, we use \textit{Mathematica} to
symbolically find the ratio 
\begin{equation}
\frac{Z_{N_{f},k}^{U(N_{c})}\left( \eta \right) }{Z_{N_{f},-k}^{U(\left\vert
k\right\vert +N_{f}-N_{c})}\left( -\eta \right) }=e^{\mathrm{i}\pi \left( 
\text{sgn}(k)\phi (N_{f},N_{c},|k|)+\varphi (N_{f},N_{c},k,m,\eta )\right) }.
\label{fullratio1}
\end{equation}

\begin{table}[tbp]
\begin{tabular}{|c|c|c|c|c|c|}
\hline
&  & \multicolumn{4}{|c|}{$k$} \\ \hline
$N_f$ & $N_c$ & -2 & -1 & 1 & 2 \\ \hline\hline
1 & 1 & $-\frac{i \left(m^2+2 m \pi \eta -2 \pi ^2 \eta ^2\right)}{2 \pi }$
& $-\frac{i \left(m^2+4 m \pi \eta -4 \pi ^2 \eta ^2\right)}{4 \pi }$ & $%
\frac{i \left(m^2-4 m \pi \eta -4 \pi ^2 \eta ^2\right)}{4 \pi }$ & $\frac{i
\left(m^2-2 m \pi \eta -2 \pi ^2 \eta ^2\right)}{2 \pi }$ \\ \hline
1 & 2 & $-\frac{i \left(m^2+2 m \pi \eta -2 \pi ^2 \eta ^2\right)}{2 \pi }$
& $-\frac{i \left(m^2+4 m \pi \eta -4 \pi ^2 \eta ^2\right)}{4 \pi }$ & $%
\frac{i \left(m^2-4 m \pi \eta -4 \pi ^2 \eta ^2\right)}{4 \pi }$ & $\frac{i
\left(m^2-2 m \pi \eta -2 \pi ^2 \eta ^2\right)}{2 \pi } $ \\ \hline
2 & 1 &  & $-\frac{i \left(m^2+4 m \pi \eta -2 \pi ^2 \eta ^2\right)}{2 \pi }
$ & $\frac{i \left(m^2-4 m \pi \eta -2 \pi ^2 \eta ^2\right)}{2 \pi }$ &  \\ 
\hline
2 & 2 & $-\frac{i \left(m^2+2 m \pi \eta -\pi ^2 \eta ^2\right)}{\pi }$ & $-%
\frac{i \left(m^2+4 m \pi \eta -2 \pi ^2 \eta ^2\right)}{2 \pi }$ & $\frac{i
\left(m^2-4 m \pi \eta -2 \pi ^2 \eta ^2\right)}{2 \pi }$ & $\frac{i
\left(m^2-2 m \pi \eta -\pi ^2 \eta^2\right)}{\pi }$ \\ \hline
\end{tabular}
\vspace{3mm}
\caption{Dependence of the phase factor of the duality on the mass and FI
terms for the theory with $N_{f}$ massive hypermultiplets, i.e. the matrix
model \eqref{Z}. For values of the parameters where the duality is not
tested, due to computer memory limitations, the cell is left empty.}
\label{tbl1.1}
\end{table}
To find this ratio we employ symbolical calculations using \textit{%
Mathematica}. Alternatively, one could attempt to numerically find %
\eqref{fullratio1} for several values of $m$ and $\eta $ and then restore
their functional dependence. The latter process is time consuming in many
aspects and we focus on the symbolic approach. As we also mentioned
previously, these calculations are very memory-demanding and one cannot
handle as many cases as presented in Table \ref{tbl1}. However for the
values of $N_{f},N_{c},k$ we compute, we do get a conclusive formula for the
function $\varphi (N_{f},N_{c},k,m,\eta )$. In particular, in Table \ref%
{tbl1.1} we present the term $\pi \varphi (N_{f},N_{c},k,m,\eta )$ in the
right hand side of \eqref{fullratio1}. Comparing the second row to the first
and the fourth row to the third one, we observe that there is no $N_{c}$
dependence in $\varphi $, hence $\varphi (N_{f},N_{c},k,m,\eta )\equiv
\varphi (N_{f},k,m,\eta )$. One can easily work out the following functional
form of $\varphi ,$ which covers all the cases presented in Table \ref%
{tbl1.1} 
\begin{eqnarray}
\varphi (N_{f},k,m,\eta ) &=&\text{sgn}(k)\left( \frac{|k|N_{f}m^{2}}{4\pi
^{2}}-\frac{\text{sgn}(k)N_{f}m\eta }{\pi }-\eta ^{2}\right)  \notag \\
&=&\frac{kN_{f}m^{2}}{4\pi ^{2}}-\frac{N_{f}m\eta }{\pi }-\text{sgn}(k)\eta
^{2}.  \label{varphi}
\end{eqnarray}%
Recalling that the mass terms in \eqref{m} are related to the mass term in %
\eqref{Z} via $2\pi m_{a}=m$ we compare the \eqref{varphi} to the phase
factors in \eqref{Z} and notice that the two expressions are almost
identical apart from a sign difference in the term $N_{f}m\eta /\pi $. 
%\textcolor{red}{Compare it with \eqref{conj2}!!!!!!!!!!!}

\begin{table}[tbp]
\begin{tabular}{|c|c|c|c|c|c|c|c|}
\hline
&  & \multicolumn{6}{|c|}{$k$} \\ \hline
$N_f$ & $N_c$ & -3 & -2 & -1 & 1 & 2 & 3 \\ \hline\hline
1 & 1 & $-\frac{3 i m^2}{4 \pi } $ & $-\frac{i m^2}{2 \pi }$ & $-\frac{i m^2%
}{4 \pi } $ & $\frac{i m^2}{4 \pi }$ & $\frac{i m^2}{2 \pi }$ & $\frac{3 i
m^2}{4 \pi }$ \\ \hline
1 & 2 &  & $-\frac{i m^2}{2 \pi }$ & $-\frac{i m^2}{4 \pi }$ & $\frac{i m^2}{%
4 \pi }$ & $\frac{i m^2}{2 \pi }$ &  \\ \hline
2 & 1 &  & $-\frac{i m^2}{\pi }$ & $-\frac{i m^2}{2 \pi }$ & $\frac{i m^2}{2
\pi }$ & $\frac{i m^2}{\pi } $ &  \\ \hline
2 & 2 & $\frac{3 im^2}{2\pi }$ & $-\frac{i m^2}{\pi }$ & $-\frac{i m^2}{2
\pi }$ & $\frac{i m^2}{2 \pi }$ & $\frac{i m^2}{\pi}$ & $\frac{3 im^2}{2\pi }
$ \\ \hline
3 & 1 &  &  & $-\frac{3 i m^2}{4 \pi }$ & $\frac{3 i m^2}{4 \pi }$ &  &  \\ 
\hline
3 & 2 &  & $-\frac{3 i m^2}{2 \pi }$ & $-\frac{3 i m^2}{4 \pi }$ & $\frac{3
i m^2}{4 \pi }$ & $\frac{3 i m^2}{2 \pi }$ &  \\ \hline
\end{tabular}
\vspace{3mm}
\caption{Dependence of the phase factor of the duality only on mass with $%
\protect\eta =0$ for the theory with $N_{f}$ massive hypermultiplets, see 
\eqref{Z}. For values of the parameters where the duality is not tested, due
to computer memory limitations, the cell is left empty.}
\label{tbl1.2}
\end{table}
While we found an expression for $\varphi $ we further test it for $\eta =0$%
, in which case we are able to explore more values of the parameters $%
N_{f},N_{c},k$. We present our findings in Table \ref{tbl1.2}, which also
provides further evidence for the functional form \eqref{varphi}.

\subsection{$N_{f}$ hypermultiplets with mass $m$ and $N_{f}$
hypermultiplets with mass $-m$}

\begin{table}[tbp]
\begin{tabular}{|c|c|c|c|c|c|c|c|c|c|c|c|}
\hline
&  & \multicolumn{10}{|c|}{$k$} \\ \hline
$N_f$ & $N_c$ & -5 & -4 & -3 & -2 & -1 & 1 & 2 & 3 & 4 & 5 \\ \hline\hline
1 & 1 & -3/4 & -1/2 & 11/12 & -1/2 & -3/4 & 3/4 & 1/2 & -11/12 & 1/2 & 3/4
\\ \hline
1 & 2 & -3/4 & 1/2 & 11/12 & 1/2 & -3/4 & 3/4 & -1/2 & -11/12 & -1/2 & 3/4
\\ \hline
1 & 3 & -3/4 & -1/2 & 11/12 & -1/2 & -3/4 & 3/4 & 1/2 & -11/12 & 1/2 & 3/4
\\ \hline
1 & 4 & -3/4 & 1/2 & 11/12 & 1/2 & NaN & NaN & -1/2 & -11/12 & -1/2 & 3/4 \\ 
\hline
2 & 1 & 3/4 & 1/2 & -7/12 & -1/2 & 3/4 & -3/4 & 1/2 & 7/12 & -1/2 & -3/4 \\ 
\hline
2 & 2 & 3/4 & -1/2 & -7/12 & 1/2 & 3/4 & -3/4 & -1/2 & 7/12 & 1/2 & -3/4 \\ 
\hline
2 & 3 & 3/4 & 1/2 & -7/12 & -1/2 & 3/4 & -3/4 & 1/2 & 7/12 & -1/2 & -3/4 \\ 
\hline
2 & 4 & 3/4 & -1/2 & -7/12 & 1/2 & 3/4 & -3/4 & -1/2 & 7/12 & 1/2 & -3/4 \\ 
\hline
2 & 5 & 3/4 & 1/2 & -7/12 & -1/2 & 3/4 & -3/4 & 1/2 & 7/12 & -1/2 & -3/4 \\ 
\hline
3 & 1 & 1/4 & -1/2 & -1/12 & -1/2 & 1/4 & -1/4 & 1/2 & 1/12 & 1/2 & -1/4 \\ 
\hline
3 & 2 & 1/4 & 1/2 & -1/12 & 1/2 & 1/4 & -1/4 & -1/2 & 1/12 & -1/2 & -1/4 \\ 
\hline
3 & 3 & 1/4 & -1/2 & -1/12 & -1/2 & 1/4 & -1/4 & 1/2 & 1/12 & 1/2 & -1/4 \\ 
\hline
3 & 4 & 1/4 & 1/2 & -1/12 & 1/2 & 1/4 & -1/4 & -1/2 & 1/12 & -1/2 & -1/4 \\ 
\hline
3 & 5 & 1/4 & -1/2 & -1/12 & -1/2 & 1/4 & -1/4 & 1/2 & 1/12 & 1/2 & -1/4 \\ 
\hline
4 & 1 & -1/4 & 1/2 & 5/12 & -1/2 & -1/4 & 1/4 & 1/2 & -5/12 & -1/2 & 1/4 \\ 
\hline
4 & 2 & -1/4 & -1/2 & 5/12 & 1/2 & -1/4 & 1/4 & -1/2 & -5/12 & 1/2 & 1/4 \\ 
\hline
4 & 3 & -1/4 & 1/2 & 5/12 & -1/2 & -1/4 & 1/4 & 1/2 & -5/12 & -1/2 & 1/4 \\ 
\hline
4 & 4 & -1/4 & -1/2 & 5/12 & 1/2 & -1/4 & 1/4 & -1/2 & -5/12 & 1/2 & 1/4 \\ 
\hline
4 & 5 & -1/4 & 1/2 & 5/12 & -1/2 & -1/4 & 1/4 & 1/2 & -5/12 & -1/2 & 1/4 \\ 
\hline
4 & 6 & -1/4 & -1/2 & 5/12 & 1/2 & -1/4 & 1/4 & -1/2 & -5/12 & 1/2 & 1/4 \\ 
\hline
5 & 1 & -3/4 & -1/2 & 11/12 & -1/2 & -3/4 & 3/4 & 1/2 & -11/12 & 1/2 & 3/4
\\ \hline
5 & 2 & -3/4 & 1/2 & 11/12 & 1/2 & -3/4 & 3/4 & -1/2 & -11/12 & -1/2 & 3/4
\\ \hline
5 & 3 & -3/4 & -1/2 & 11/12 & -1/2 & -3/4 & 3/4 & 1/2 & -11/12 & 1/2 & 3/4
\\ \hline
5 & 4 & -3/4 & 1/2 & 11/12 & 1/2 & -3/4 & 3/4 & -1/2 & -11/12 & -1/2 & 3/4
\\ \hline
5 & 5 & -3/4 & -1/2 & 11/12 & -1/2 & -3/4 & 3/4 & 1/2 & -11/12 & 1/2 & 3/4
\\ \hline
5 & 6 & -3/4 & 1/2 & 11/12 & 1/2 & -3/4 & 3/4 & -1/2 & -11/12 & -1/2 & 3/4
\\ \hline
5 & 7 & -3/4 & -1/2 & 11/12 & -1/2 & -3/4 & 3/4 & 1/2 & -11/12 & 1/2 & 3/4
\\ \hline
\end{tabular}
\vspace{3mm}
\caption{Values of the parameter $\protect\theta $ for different values of $%
N_{f},N_{c}$ and $k$ of the theory with $2N_{f}$ massless hypermultiplets,
actually the massless case of the matrix model \eqref{Zm2}. NaN refers to
instances where the dual theory is not well defined because the dual number
of colours is negative.}
\label{tbl2}
\end{table}
Next we present the results on the phase factor for the theory with $N_{f}$
hypermultiplets of mass $m$ and $N_{f}$ hypermultiplets of mass $-m$
discussed in Section \ref{2Nf_Hypermultiplets}. In the case of $2N_{f}$
hypermultiplets, Giveon-Kutasov duality is between $U(N_{c})$ and $%
U(|k|+2N_{f}-N_{c})$, where $k$ is the Chern-Simons level. Similarly to the
previous Section, we implement in \textit{Mathematica} the solution %
\eqref{ZNf} with \eqref{fifj} and proceed in two steps. First we numerically
and/or symbolically compute the phase factor for the massless and $\eta =0$
case. Then we turn on the mass terms and compute the mass dependence of the
phase. For the first step, we numerically compute the ratio 
\begin{equation}
\frac{\widetilde{Z}_{N_{f},k}^{U(N_{c})}\left( 0\right) }{\widetilde{Z}%
_{N_{f},-k}^{U(\left\vert k\right\vert +2N_{f}-N_{c})}\left( 0\right) }=e^{%
\mathrm{i}\pi \theta }
\end{equation}%
for $0<|k|\leq 5$. We present the values of the parameter $\theta $ in Table %
\ref{tbl2}. One may proceed as before to find a universal expression as a
function of $N_{f},N_{c},k$ that covers all the values in the table. However
we observe that the values of $\theta$ for $N_{f}=1,2,3,4$ in Table \ref%
{tbl2} are identical to the values for $N_{f}=2,4,6,8$ in Table \ref{tbl1}.
This is expected, because in the massless case the matrix models \eqref{Z}
and \eqref{Zm2} are identical with $N_f$ replaced by $2N_f$. Since we
already have an expression for the quadratic function of the former matrix
model, we use \eqref{QuadraticFunction} replacing $N_{f}$ with $2N_{f}$ and
further test the remaining values of $N_{f}$ and $N_{c}$, confirming that $%
\phi \left( 2N_{f},N_{c},k\right) $ does give the expected results, which
means that $e^{\mathrm{i}\pi \phi \left( 2N_{f},N_{c},k\right) }/e^{\mathrm{i%
}\pi \theta }=1$.

\begin{table}[tbp]
\begin{tabular}{|c|c|c|c|c|c|c|c|}
\hline
&  & \multicolumn{6}{|c|}{$k$} \\ \hline
$N_f$ & $N_c$ & -3 & -2 & -1 & 1 & 2 & 3 \\ \hline\hline
1 & 1 & $-\frac{3 i m^2}{2 \pi }$ & $-\frac{i m^2}{\pi }$ & $-\frac{i m^2}{2
\pi } $ & $\frac{i m^2}{2 \pi }$ & $\frac{i m^2}{\pi }$ & $\frac{3 i m^2}{2
\pi }$ \\ \hline
1 & 2 & $-\frac{3 i m^2}{2 \pi }$ & $-\frac{i m^2}{\pi }$ & $-\frac{i m^2}{2
\pi }$ & $\frac{i m^2}{2 \pi }$ & $\frac{i m^2}{\pi }$ & $\frac{3 i m^2}{2
\pi }$ \\ \hline
2 & 1 &  &  & $-\frac{i m^2}{\pi }$ &  &  &  \\ \hline
2 & 2 &  &  & $-\frac{i m^2}{\pi }$ & $\frac{i m^2}{\pi }$ &  &  \\ \hline
2 & 3 &  & $-\frac{2 i m^2}{\pi }$ & $-\frac{i m^2}{\pi }$ & $\frac{i m^2}{%
\pi }$ & $\frac{2 i m^2}{\pi }$ &  \\ \hline
\end{tabular}
\vspace{3mm}
\caption{Dependence of the phase factor of the duality on mass for the
theory with $N_{f}$ hypermultiplets with masses $m$ and $N_{f}$
hypermultiplets with masses $-m$. For values of the parameters where the
duality is not tested, due to computer memory limitations, the cell is left
empty.}
\label{tbl2.2}
\end{table}
Having determined the quadratic $k$-dependence of the phase factor, we turn
on the masses. We present the results for several cases of the parameters $%
N_{f},N_{c},k$ in Table \ref{tbl2.2}. Comparing with the results in Table %
\ref{tbl1.2} we observe that they differ by a factor of 2, due to the fact
that we now have two copies of $N_{f}$ hypermultiplets. Therefore one may
safely assume that \eqref{varphi} is still valid for $\eta =0$ and $N_{f}$
replaced by $2N_{f}$.

The case of non-zero FI term will be examined in a more general setting of
the theory, with $N_{f}$ hypermultiplets of mass $m_{1}$ and $N_{f}$
hypermultiplets of mass $m_{2}$, which is the topic of the next Section.

\subsection{$N_{f}$ hypermultiplets with mass $m_{1}$ and $N_{f}$
hypermultiplets with mass $m_{2}$}

In this section we study a theory similar to \eqref{Zm2} but with masses $%
m_{1}$ and $m_{2}$ and a non-zero FI term, $\eta $, 
\begin{equation}
\widehat{Z}_{N_{f}}^{U(N)}=\frac{1}{\left( 2\pi \right) ^{N}N!}\int {%
d^{N}\!\mu }\frac{\prod_{i<j}4\sinh ^{2}(\frac{1}{2}(\mu _{i}-\mu _{j}))\
e^{-\frac{1}{2g}\sum_{i}\mu _{i}^{2}+\mathrm{i}\eta \sum_{i}\mu _{i}}}{%
\prod_{i}\left( 4\cosh (\frac{1}{2}(\mu _{i}+m_{1}))\cosh (\frac{1}{2}(\mu
_{i}+m_{2}))\right) ^{N_{f}}}.  \label{Zm1m2}
\end{equation}%
We solve it by substituting $z=ce^{\mu }$ and $c=e^{g(N-N_{f})}$ and
following the steps in Sections \ref{NfGeneric} and \ref{2Nf_Hypermultiplets}%
. We find 
\begin{equation}
\widehat{Z}_{N_{f}}^{U(N)}=e^{\frac{1}{2}(m_{1}+m_{2})NN_{f}}c^{-\frac{N}{2}%
(N+N_{f}+2\mathrm{i}\eta )}\det \left( \left( h_{i},h_{j}\right) \right)
_{i,j=0}^{N-1},
\end{equation}%
where the elements of the $(h_{i},h_{j})$ matrix are given by 
\begin{eqnarray}
(h_{i},h_{j}) &=&\frac{c^{i+j+1+\mathrm{i}\eta }e^{-\frac{1}{2g}\left( \ln
c\right) ^{2}}}{\left( e^{m_{1}}-e^{m_{2}}\right) ^{N_{f}}}%
\sum_{s=0}^{N_{f}-1}\frac{\binom{N_{f}+s-1}{s}}{(N_{f}-s-1)!}\left( \frac{%
e^{m_{1}+m_{2}}}{e^{m_{1}}-e^{m_{2}}}\right) ^{s}\times   \notag \\
&&\sum_{n=0}^{N_{f}-s-1}\left( (-1)^{N_{f}}C_{N_{f}-s-1,n}I^{(n)}(\hat{\ell}%
+s,m_{1})+(-1)^{s}C_{N_{f}-s-1,n}I^{(n)}(\hat{\ell}+s,m_{2})\right) 
\label{hihj}
\end{eqnarray}%
where $\hat{\ell}=i+j+1-N+\mathrm{i}\eta $ and $C_{p,q}$ given by \eqref{Cpq}%
. One can verify that for $m_{1}=-m_{2}=m$ and $\eta =0$ finds the solution %
\eqref{fifj}.

We again implement \eqref{hihj} in \textit{Mathematica} and test the
Giveon-Kutasov duality for low values of the parameters $N_{f},N_{c},k$. 
%Due to the computational cost we are not able to verify the duality for both $m\neq$ and $\eta \neq 0$. Instead we test it when either of the parameters is zero. 
For $N_{f}=N_{c}=k=1$ we get 
\begin{equation}
\frac{\widehat{Z}_{N_{f}}^{U(N_{c})}}{\widehat{Z}%
_{N_{f}}^{U(|k|+2N_{f}-N_{c})}}=e^{\frac{3\pi \mathrm{i}}{4}}e^{\frac{%
\mathrm{i}\left( m_{1}^{2}+m_{2}^{2}\right) }{4\pi }-\mathrm{i}\eta
(m_{1}+m_{2})-\mathrm{i}\pi \eta ^{2}}.
\end{equation}%
For $N_{f}=1,N_{c}=2,k=1$ we find 
\begin{equation}
\frac{\widehat{Z}_{N_{f}}^{U(N_{c})}}{\widehat{Z}%
_{N_{f}}^{U(|k|+2N_{f}-N_{c})}}=e^{\frac{3\pi \mathrm{i}}{4}}e^{\frac{%
\mathrm{i}\left( m_{1}^{2}+m_{2}^{2}\right) }{4\pi }-\mathrm{i}%
(m_{1}+m_{2})\eta -\mathrm{i}\pi \eta ^{2}},
\end{equation}%
whereas for $N_{f}=1,N_{c}=2,k=-2$ the ratio becomes 
\begin{equation}
\frac{\widehat{Z}_{N_{f}}^{U(N_{c})}}{\widehat{Z}%
_{N_{f}}^{U(|k|+2N_{f}-N_{c})}}=e^{\frac{\pi \mathrm{i}}{2}}e^{-\frac{%
\mathrm{i}\left( m_{1}^{2}+m_{2}^{2}\right) }{2\pi }-\mathrm{i}%
(m_{1}+m_{2})\eta +\mathrm{i}\pi \eta ^{2}}.
\end{equation}%
From these few examples we do observe a pattern, as we notice that only the $%
k$-quadratic phase depends on $N_{c}$ and not the mass and FI terms. The
latter terms are validated through 
\begin{equation}
\widehat{\varphi }(N_{f},k,m_{1},m_{2},\eta )=\frac{%
kN_{f}(m_{1}^{2}+m_{2}^{2})}{4\pi ^{2}}-\frac{N_{f}(m_{1}+m_{2})\eta }{\pi }-%
\text{sgn}(k)\eta ^{2},
\end{equation}%
which is a generalization of \eqref{varphi} and is in agreement with the
phase in \eqref{conj2}, again up to a sign difference in the $%
(m_{1}+m_{2})\eta $ term, as mentioned above.

\subsection*{Acknowledgements}

We thank David Garc\'{\i}a, Luis Melgar, Jorge Russo and Christian Vergu for
comments and discussion. The work of MT is funded by an Investigador FCT
position at Universidade de Lisboa (Reference IF/01767/2014). MT
acknowledges former support at Universidad Complutense de Madrid, from
MINECO (grant MTM2011-26912), Comunidad de Madrid (grant QUITEMAD+-CM, ref.
S2013/ICE-2801) and the European CHIST-ERA project CQC (funded partially by
MINECO grant PRI-PIMCHI-2011-1071).

\end{document}